\documentclass[pdflatex,sn-mathphys-ay]{sn-jnl}

\usepackage{graphicx}%
\usepackage{multirow}%
\usepackage{amsmath,amssymb,amsfonts}%
\usepackage{amsthm}%
\usepackage{mathrsfs}%
\usepackage[title]{appendix}%
\usepackage{xcolor}%
\usepackage{textcomp}%
\usepackage{manyfoot}%
\usepackage{booktabs}%
\usepackage{listings}%
\usepackage{booktabs}
\usepackage{array}
\usepackage{algorithm}
\usepackage{subcaption}
\usepackage{algorithmic}
\usepackage{multirow}
\usepackage{braket}
\usepackage{threeparttable}

\theoremstyle{thmstyleone}%
\theoremstyle{thmstyletwo}%

\theoremstyle{thmstylethree}%

\begin{document}

\title[Article Title]{Q-AGNN: Quantum-Enhanced Attentive Graph Neural Network for Intrusion Detection}


\author*[1]{\fnm{Devashish} \sur{Chaudhary}}\email{s224281473@deakin.edu.au}

\author[1]{\fnm{Sutharshan} \sur{Rajasegarar}}\email{sutharshan.rajasegarar@deakin.edu.au}
\equalcont{These authors contributed equally to this work.}

\author[1]{\fnm{Shiva Raj} \sur{Pokhrel}}\email{shiva.pokhrel@deakin.edu.au}
\equalcont{These authors contributed equally to this work.}

\affil[1]{\orgdiv{School of Information Technology}, 
\orgname{Deakin University}, 
\orgaddress{\city{Geelong}, \state{Victoria}, \postcode{3220}, \country{Australia}}}


\abstract{With the rapid growth of interconnected devices, accurately detecting malicious activities in network traffic has become increasingly challenging. Most existing deep learning-based intrusion detection systems treat network flows as independent instances, thereby failing to exploit the relational dependencies inherent in network communications. To address this limitation, we propose Q-AGNN, a Quantum-Enhanced Attentive Graph Neural Network for intrusion detection, where network flows are modeled as nodes and edges represent similarity relationships. Q-AGNN leverages parameterized quantum circuits (PQCs) to encode multi-hop neighborhood information into a high-dimensional latent space, inducing a bounded quantum feature map that implements a second-order polynomial graph filter in a quantum-induced Hilbert space. An attention mechanism is subsequently applied to adaptively weight the quantum-enhanced embeddings, allowing the model to focus on the most influential nodes contributing to anomalous behavior. Extensive experiments conducted on four benchmark intrusion detection datasets demonstrate that Q-AGNN achieves competitive or superior detection performance compared to state-of-the-art graph-based methods, while consistently maintaining low false positive rates under hardware-calibrated noise conditions. Moreover, we also executed the Q-AGNN framework on actual IBM quantum hardware to demonstrate the practical operability of the proposed pipeline under real NISQ conditions. These results highlight the effectiveness of integrating quantum-enhanced representations with attention mechanisms for graph-based intrusion detection and underscore the potential of hybrid quantum-classical learning frameworks in cybersecurity applications.}




\maketitle
\section{Introduction}

The rapid proliferation of interconnected devices, cloud services, and Internet-of-Things (IoT) infrastructures has significantly increased the complexity and scale of modern networks(\cite{shahraki2020survey}). While this connectivity enables efficient data exchange and intelligent services, it also expands the attack surface, making networks increasingly vulnerable to sophisticated cyber threats. Intrusion Detection Systems (IDS) play a crucial role in identifying malicious activities and safeguarding network integrity. However, the evolving nature of attacks and the high volume of network traffic pose substantial challenges to existing IDS solutions(\cite{neto2025deep}).

Traditional machine learning and deep learning-based intrusion detection approaches typically model network traffic flows as independent samples, relying on handcrafted or learned feature representations(\cite{rajendran2019detection}). Although such methods have achieved promising performance, they often overlook the inherent relational dependencies among network entities, such as communication patterns, temporal correlations, and shared behavioral characteristics. Ignoring these dependencies limits the ability of IDS models to detect complex and coordinated attacks that manifest across multiple network flows(\cite{zhong2024survey}).

To address this limitation, recent studies have adopted graph-based learning techniques, where network traffic is represented as graphs and detection is performed using Graph Neural Networks (GNNs)(\cite{lo2021graphsage,caville2022anomal,altaf2023new,altaf2023ne, jiang2025orthrus, bilot2025sometimes}). While these methods improve detection by aggregating neighborhood information, they remain constrained by the expressiveness of classical aggregation functions and often struggle to capture high-order, non-linear relationships in large-scale or noisy network environments. Moreover, deeper GNN architectures may suffer from issues such as over-smoothing and performance degradation(\cite{li2024another}).

In parallel, quantum machine learning (QML) has emerged as a promising paradigm capable of encoding complex, high-dimensional data using quantum feature spaces(\cite{hdaib2024quantum, sakhnenko2022hybrid}). Parameterized quantum circuits (PQCs) have been shown to provide expressive representations that may enhance learning performance in certain tasks. Nevertheless, the practical deployment of fully quantum models is hindered by current hardware limitations, including noise and restricted qubit counts, motivating the development of hybrid quantum-classical architectures that can operate effectively in the near-term noisy intermediate-scale quantum (NISQ) era(\cite{ahmed2025comparative}).

Motivated by these observations, we propose Q-AGNN, a Quantum-Enhanced Attentive Graph Neural Network for intrusion detection. In Q-AGNN, network flows are modeled as nodes and edges encode similarity relationships. Multi-hop neighborhood features are first encoded using parameterized quantum circuits to capture complex correlations, after which an attention mechanism selectively emphasizes the most influential quantum-enhanced embeddings. The resulting node representations are then used for accurate intrusion classification. The main contributions of this work are summarized as follows:

\begin{itemize}
    \item We introduce a novel hybrid quantum-classical graph-based intrusion detection framework that integrates PQCs with attention mechanisms.
    \item We design a quantum-enhanced neighborhood encoding strategy that captures multi-hop relational information in network graphs. We implement and analyze the impact of quantum noise and highlight the robustness of the proposed approach under realistic NISQ settings.
    \item We conduct extensive experiments on four benchmark intrusion detection datasets, demonstrating that Q-AGNN achieves competitive or superior performance compared to state-of-the-art GNN-based methods while minimizing false positive rates.
   
    \item We train and evaluate Q-AGNN directly on actual IBM quantum hardware using a feasible subset of real network flows, providing concrete experimental evidence of the practical viability of the proposed approach under realistic quantum noise. These results establish an important stepping stone toward deployment on fault-tolerant quantum systems. 
\end{itemize}

To the best of our knowledge, this study represents one of the earliest demonstrations of a quantum-enhanced graph neural network with attention mechanisms applied to intrusion detection in noisy quantum hardware environments.

\section{Preliminaries and Related Work}

This section provides a review of relevant works at the intersection of graph-based intrusion detection and quantum machine learning, as summarized in Table~\ref{tab:comparison} and positions the proposed Q-AGNN architecture within the landscape.

GNNs have emerged as a dominant paradigm for modeling relational structure in network traffic, consistently outperforming tabular and sequence-based methods in intrusion detection tasks. Early work demonstrated that explicitly encoding network topology enables more accurate detection of coordinated and distributed attacks. \cite{cao2021detecting} proposed the Spatial-Temporal Graph Convolutional Network (ST-GCN), which models SDN switches as graph nodes and captures both structural and temporal dependencies, achieving significant improvements in DDoS detection.

Subsequent research has moved beyond static, topology-driven graphs toward feature-driven graph construction. \cite{hosler2024graph} modeled individual network flows as nodes connected via feature similarity, enabling improved detection of stealthy and evasive attacks that do not manifest through fixed IP or port-based relationships. In parallel, \citet{lo2021graphsage} proposed E-GraphSAGE, which incorporates edge attributes into neighborhood aggregation and demonstrates that relational traffic features play a critical role in learning discriminative embeddings. Attention-based aggregation mechanisms further advanced this direction, as exemplified by the Multi-Graph GNN~\cite{altaf2023new}, which enables adaptive weighting of neighbors through a hybrid spectral-spatial architecture.

In parallel, QML has gained attention as a means of capturing complex, non-linear patterns in network traffic. Prior studies have explored quantum autoencoders, quantum support vector classifiers, and hybrid quantum-classical generative models for anomaly and intrusion detection, demonstrating advantages in expressivity and robustness under certain conditions(\cite{hdaib2024quantum,kumar2025quids,rahman2023quantum}). Frameworks such as QuantumNetSec(\cite{abreu2025quantumnetsec}) further showed that carefully designed variational quantum circuits can achieve competitive performance on noisy intermediate-scale quantum (NISQ) devices.

More recently, efforts have begun to combine graph learning with quantum computation. Quantum graph neural networks extend classical message passing into the quantum (\cite{verdon2019quantum,zheng2021quantum,devale2025design,innan2024financial}), while QEGraphSAGE(\cite{hoang2025intrusion}) integrates variational quantum circuits into edge-aware graph learning for IoT traffic analysis. Closest to this work, \cite{ning2025quantum} proposed the Quantum Graph Attention Network (QGAT), which replaces classical attention with variational quantum circuits to generate multiple attention coefficients via quantum parallelism.

Building on these advances, we propose \textbf{Q-AGNN}, a hybrid quantum-classical graph architecture that uniquely combines quantum-enhanced node feature encoding with multi-hop attention-based neighborhood aggregation for flow-level intrusion detection. Unlike prior quantum GNNs, Q-AGNN explicitly targets realistic intrusion detection settings and is evaluated under practical quantum noise constraints.


\begin{table}[t]
\caption{Comparison of representative graph-based and quantum-enhanced intrusion detection approaches.}
\label{tab:comparison}
\centering
\setlength{\tabcolsep}{3.6pt}
\renewcommand{\arraystretch}{1.1}

\begin{tabular*}{\textwidth}{@{\extracolsep\fill}lccccc p{2.1cm}}
\toprule
Method & Graph & Attn\footnotemark[1] & Q-Enc\footnotemark[1] & Hybrid & Noise & Task \\
\midrule
ST-GCN & Topology & No & No & No & N/A & DDoS \\
E-GraphSAGE & Feature & No & No & No & N/A & IDS \\
Multi-Graph GNN & Multi-view & Yes & No & No & N/A & IDS \\
QuantumNetSec & None & No & Yes & Yes & Yes & IDS \\
QEGraphSAGE & Feature & No & Yes & Yes & Limited & IoT IDS \\
QGAT & Feature & Quantum & Yes & Yes & Limited & Generic Graph \\
Q-AGNN (Ours) & Feature & Multi-hop & Yes & Yes & Yes & IDS \\
\bottomrule
\end{tabular*}

\footnotetext[1]{Attn: Attention; Q-Enc: Quantum Encoding.}
\end{table}

\section{Proposed Methodology}

In the standard formulation, Graph Convolutional Networks (GCNs)(\cite{kipf2016semi}) assume that all neighboring nodes contribute equally to the representation of a target node. The layer-wise propagation rule of a GCN is
\begin{equation}
{H}^{(l+1)} =
\sigma\!\left(
\hat{{D}}^{-\frac{1}{2}}
\hat{{A}}
\hat{{D}}^{-\frac{1}{2}}
{H}^{(l)}
{W}^{(l)}
\right),
\end{equation}
where $\hat{{A}} = {A} + {I}$, with ${A} \in \{0,1\}^{N \times N}$ being the adjacency matrix of the graph (where $A_{ij}=1$ if node $i$ is connected to node $j$, and $0$ otherwise) and ${I}$ denoting the identity matrix, which adds self-loops to all nodes; $\hat{{D}}$ is the corresponding degree matrix; ${H}^{(l)}$ and ${H}^{(l+1)}$ denote the node feature matrices at layers $l$ and $l+1$, respectively; ${W}^{(l)}$ is a trainable weight matrix; and $\sigma(\cdot)$ denotes a nonlinear activation function. This normalization-based aggregation treats all neighbors uniformly, which may be suboptimal in many real-world applications.

\begin{figure}[t]
    \centering
    \includegraphics[width=1.00\linewidth]{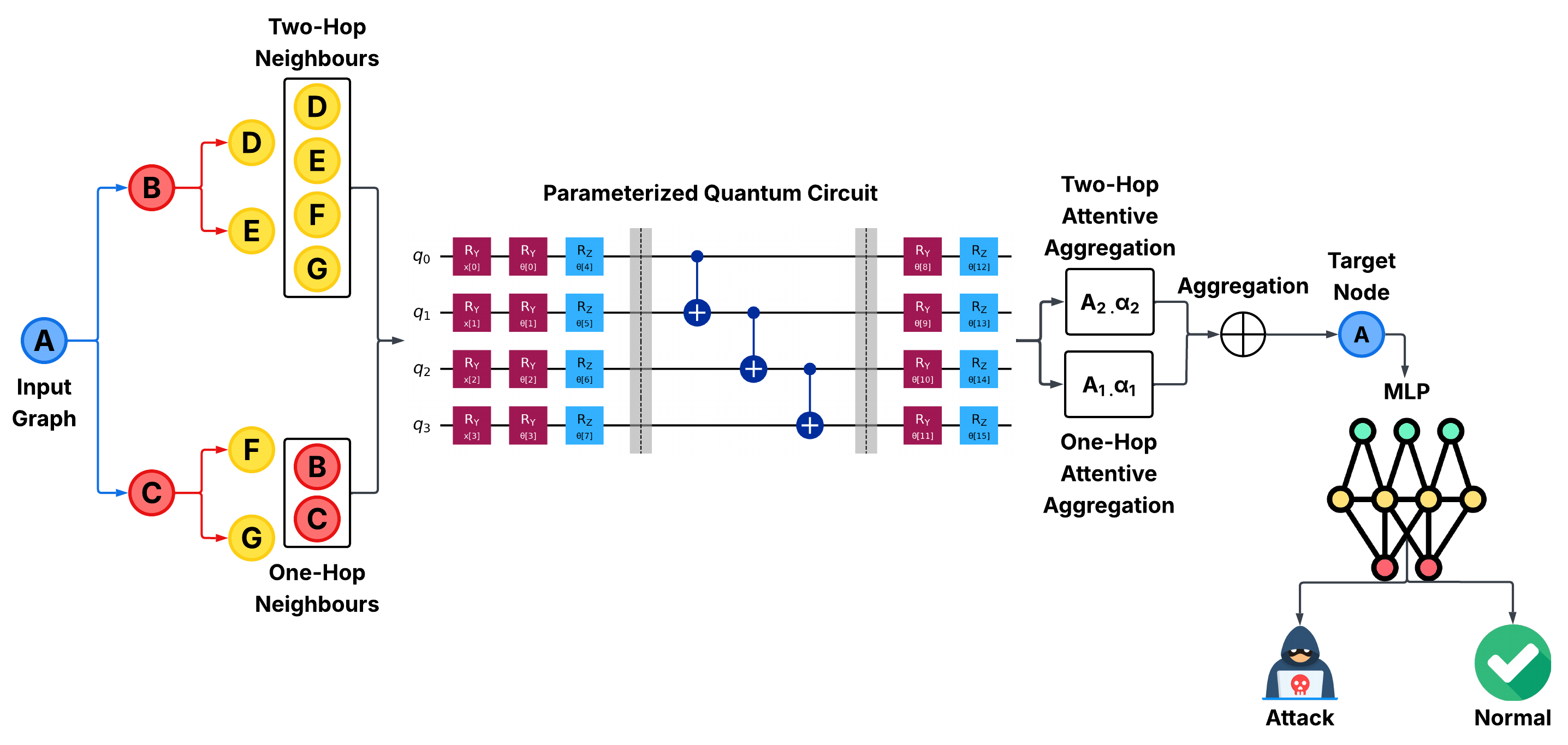}
    \caption{Overview of the proposed Q-AGNN architecture. Node features are encoded into quantum states via angle encoding and transformed using parameterized quantum circuits. Quantum-enhanced embeddings are aggregated using one-hop and two-hop attention mechanisms, followed by a classical MLP for intrusion detection.}
    \label{fig:architecture}
\end{figure}

To address scalability and generalization limitations of GCNs, GraphSAGE was proposed as an inductive framework that learns aggregation functions over sampled neighborhoods rather than relying on the full graph structure(\cite{hamilton2017inductive}). In GraphSAGE, the representation of node $i$ is updated by first aggregating its neighbors’ features using a predefined function and then combining this aggregated representation with the node’s own features, expressed as
\begin{equation}
{h}_i' =
\sigma\!\left(
{W}
\left[
{h}_i \;\Vert\;
\text{AGG}\!\left(
\left\{
{h}_j : j \in \mathcal{N}(i)
\right\}
\right)
\right]
\right),
\end{equation}
where ${h}_i$ and ${h}_i'$ denote the embedding of node $i$ before and after aggregation, respectively; $\text{AGG}(\cdot)$ is a predefined aggregation function such as mean or pooling; and ${W}$ is a trainable weight matrix.

While GraphSAGE enables inductive learning on large or evolving graphs, its aggregation strategy treats all neighbors equally, which can dilute critical signals in security-sensitive applications such as IDS. In graph-based IDS, nodes may represent hosts or network flows, and edges represent communication relationships. Some neighbors correspond to benign traffic, while others may exhibit anomalous behavior such as abnormal packet rates, bursty communication, or irregular timing patterns. Fixed aggregation functions are unable to explicitly distinguish between such heterogeneous interactions.

This limitation motivates the use of Graph Attention Networks (GATs)(\cite{velivckovic2017graph}), which introduce a self-attention mechanism to adaptively weight neighbor contributions. In GATs, the updated feature representation of node $i$ is computed as
\begin{equation}
{h}_i' =
\sigma\left(
\sum_{j \in \mathcal{N}(i)} \alpha_{ij} {W} {h}_j
\right),
\end{equation}
where the attention coefficients $\alpha_{ij}$ reflect the importance of neighbor $j$ to node $i$ and are obtained using a shared attention mechanism:
\begin{equation}
\alpha_{ij} =
\frac{
\exp\left(\text{LeakyReLU}\left({a}^{\top} [{W} {h}_i \, \| \, {W} {h}_j]\right)\right)
}{
\sum_{k \in \mathcal{N}(i)} \exp\left(\text{LeakyReLU}\left({a}^{\top} [{W} {h}_i \, \| \, {W} {h}_k]\right)\right)
}.
\end{equation}
By learning data-dependent attention weights, graph attention networks (GATs) selectively amplify security-critical neighborhood interactions while attenuating redundant or weakly informative connections, making them particularly effective for graph-based intrusion detection.

Building on these classical GNN principles, we propose \textbf{Q-AGNN}, a hybrid quantum-classical graph architecture that integrates parameterized quantum circuits (PQCs) with explicit multi-hop attention-based aggregation to learn expressive node representations for intrusion detection. Fig.~\ref{fig:architecture} presents an overview of the Q-AGNN framework, while Algorithm~\ref{alg:qagnn} details the end-to-end training and inference procedure.

Each node feature vector ${x}_i \in \mathbb{R}^{F}$ is first mapped to a quantum state using angle encoding:
\begin{equation}
|\psi_i\rangle = \bigotimes_{k=1}^{F} R_Y(x_{ik})|0\rangle,
\end{equation}
where $R_Y(\cdot)$ denotes a rotation about the $Y$ axis parameterized by feature $x_{ik}$.

The encoded state is then transformed by a parameterized EfficientSU2 ansatz with linear entangling topology. Each layer $\ell = 1, \dots, L$ of the ansatz applies pre-entangling single-qubit rotations $R_Y$ and $R_Z$ on all qubits, followed by linear CNOT gates connecting qubit $q$ to $q+1$, and then post-entangling rotations. One layer is expressed as:
\begin{align}
U_{\text{EffSU2}}^{(\ell)} &=
\underbrace{\bigotimes_{q=1}^{F} R_Y(\theta_q^{(\ell,1)}) R_Z(\theta_q^{(\ell,2)})}_{\text{pre-entangling}} \nonumber \\
&\quad \underbrace{\prod_{q=1}^{F-1} \text{CNOT}_{q,q+1}}_{\text{entanglement}} \nonumber \\
&\quad \underbrace{\bigotimes_{q=1}^{F} R_Y(\theta_q^{(\ell,3)}) R_Z(\theta_q^{(\ell,4)})}_{\text{post-entangling}}.
\end{align}

The final quantum state after $L$ layers is
\begin{equation}
|\phi_i(\boldsymbol{\theta})\rangle =
U_{\text{EffSU2}}^{(L)} \cdots U_{\text{EffSU2}}^{(1)} |\psi_i\rangle,
\end{equation}
where $\boldsymbol{\theta}$ denotes all trainable parameters.

A classical embedding is obtained via expectation values of observables:
\begin{equation}
z_i = \langle \phi_i(\boldsymbol{\theta}) \,|\, \hat{O} \,|\, \phi_i(\boldsymbol{\theta}) \rangle,
\end{equation}
where $z_i$ is the PQC-generated embedding of node $i$.

Rather than iterative message passing, Q-AGNN performs explicit multi-hop aggregation through powers of the adjacency matrix. Let $A$ be the adjacency matrix of the cosine-similarity graph. One-hop and two-hop interactions are obtained as
\begin{equation}
A^{(1)} = A, \qquad A^{(2)} = A^2.
\end{equation}

The aggregated embeddings are computed as polynomial graph filters:
\begin{equation}
    z_i^{(1)} = \sum_{j \in V} A^{(1)}_{ij}\,\alpha^{(1)}_{j}\, z_j,
\end{equation}

\begin{equation}
    z_i^{(2)} = \sum_{j \in V} A^{(2)}_{ij}\,\alpha^{(2)}_{j}\, z_j,
\end{equation}

where $z_j$ is the PQC embedding of neighboring node $j$. The attention weights $\alpha^{(1)}$ and $\alpha^{(2)}$ are computed at the node level and normalized globally across the graph, emphasizing informative nodes during aggregation.

The final node representation combines original and aggregated embeddings:
\begin{equation}
h_i' = \sigma \left( z_i + z_i^{(1)} + z_i^{(2)} \right),
\end{equation}
where $\sigma(\cdot)$ is a nonlinear activation.

A multilayer perceptron produces the intrusion score:
\begin{equation}
y_i = \mathrm{MLP}(h_i').
\end{equation}

\begin{algorithm}[t]
\caption{Q-AGNN}
\label{alg:qagnn}
\begin{algorithmic}[1]

\REQUIRE Graph $G=(V,E)$ with adjacency matrix $\mathbf{A}$, node features $\{\mathbf{h}_i \in \mathbb{R}^F\}_{i \in V}$, quantum parameters $\boldsymbol{\theta}$, attention parameters $\boldsymbol{\alpha}^{(1)}, \boldsymbol{\alpha}^{(2)}$

\ENSURE Node-level intrusion scores $\{{y}_i\}_{i \in V}$

\textbf{Quantum Embedding Generation (Node-wise)}
\FOR{each node $i \in V$}
    \STATE Encode node features:
    $|\psi_i\rangle = \bigotimes_{k=1}^{F} R_Y(h_{ik})|0\rangle$
    \STATE Apply parameterized circuit:
    $|\phi_i\rangle = U(\boldsymbol{\theta})|\psi_i\rangle$
    \STATE Measure observables to obtain embedding:
    ${z}_i = \langle \phi_i | \hat{O} | \phi_i \rangle$
\ENDFOR

\textbf{Polynomial Multi-Hop Aggregation}
\STATE Compute one-hop adjacency $\mathbf{A}^{(1)} = \mathbf{A}$
\STATE Compute two-hop adjacency $\mathbf{A}^{(2)} = \mathbf{A}^2$

\FOR{each node $i \in V$}
    \STATE One-hop aggregation:${z}_i^{(1)} = \sum_{j \in V} A^{(1)}_{ij}\,\alpha^{(1)}_{j}\,{z}_j$
    
    \STATE Two-hop aggregation:${z}_i^{(2)} = \sum_{j \in V} A^{(2)}_{ij}\,\alpha^{(2)}_{j}\,{z}_j$

\ENDFOR

\textbf{Fusion and Prediction}
\FOR{each node $i \in V$}
    \STATE Fuse embeddings:
    \[
    {h}_i' \leftarrow \sigma\!\left({z}_i + {z}_i^{(1)} + {z}_i^{(2)}\right)
    \]
    \STATE Predict intrusion score:
    ${y}_i \leftarrow \mathrm{MLP}({h}_i')$
\ENDFOR

\RETURN $\{{y}_i\}_{i \in V}$

\end{algorithmic}
\end{algorithm}

This formulation reveals that Q-AGNN implements a second-order polynomial graph filter over quantum-enhanced node embeddings, decoupling nonlinear feature transformation in the quantum domain from efficient relational aggregation in the classical domain. Algorithm~\ref{alg:qagnn} summarizes the forward pass of the proposed Q-AGNN model. The hybrid quantum-classical design of Q-AGNN is motivated by the complementary strengths of quantum computation and classical graph learning. Parameterized EfficientSU2 circuits induce highly nonlinear transformations in a high-dimensional Hilbert space, where quantum superposition enables compact representation of multiple feature interactions and entanglement captures higher-order correlations. These properties are well suited to intrusion detection, in which malicious behavior often manifests through subtle, non-linear dependencies among traffic statistics, temporal dynamics, and protocol attributes.

At the same time, current quantum hardware operates in the NISQ regime(\cite{preskill2018quantum}), with limited qubit counts, shallow circuit depths, and non-negligible noise. Fully quantum graph learning pipelines are therefore not yet practical. Q-AGNN adopts a hybrid architecture in which PQCs serve as expressive feature encoders, while classical attention-based aggregation and prediction layers provide scalability, numerical stability, and efficient optimization. This design supports end-to-end training on classical hardware, noisy simulators, and near-term quantum processors without assuming fault-tolerant quantum computation. From a longer-term perspective, Q-AGNN represents a transitional architecture toward fully quantum graph learning. As scalable fault-tolerant quantum systems mature, increasingly larger components of the pipeline, including neighborhood aggregation and attention computation, may be migrated to the quantum domain. In such settings, quantum superposition could enable parallel exploration of multiple neighborhood configurations, while entanglement may naturally encode multi-hop relational structure without explicit iterative message passing. The proposed framework is therefore both practically viable in the near term and aligned with the anticipated evolution of quantum computing.

\subsection{Design Considerations}

The architectural design of Q-AGNN is grounded in established principles from quantum feature maps and graph signal processing, with the objective of achieving high representational expressivity while ensuring numerical stability and efficient optimization under realistic computational constraints.

Let $\mathbf{x}_v \in \mathbb{R}^d$ denote the classical feature vector associated with node $v$. Each node feature is encoded by a parameterized quantum circuit (PQC) $U(\mathbf{x}_v, \boldsymbol{\theta})$ acting on $n$ qubits, producing a quantum state
\begin{equation}
    \ket{\psi(\mathbf{x}_v)} = U(\mathbf{x}_v, \boldsymbol{\theta}) \ket{0}^{\otimes n}.
\end{equation}

Node embeddings are obtained by measuring expectation values of Pauli observables $\{P_k\}$,
\begin{equation}
    z_{v,k} = \langle \psi(\mathbf{x}_v) | P_k | \psi(\mathbf{x}_v) \rangle,
\end{equation}

yielding a bounded embedding $z_{v,k} \in [-1,1]$ for all $v$ and $k$. This boundedness is an intrinsic property of quantum expectation values and ensures that the resulting feature representations remain uniformly bounded when composed with classical neural layers. Consequently, Q-AGNN avoids uncontrolled activation growth and supports numerically stable gradient-based optimization.

Rather than employing deep iterative message passing, Q-AGNN performs explicit aggregation over first- and second-order neighborhoods using powers of the adjacency matrix. Let $A \in \mathbb{R}^{N \times N}$ denote the adjacency matrix of the cosine-similarity graph constructed over network flows. The aggregation operator is defined over $A$ and $A^2$, corresponding to one-hop and two-hop neighborhoods, respectively. This formulation aligns with polynomial graph filtering approaches in graph signal processing, where higher-order structural information is captured through low-degree polynomials of the graph shift operator rather than deep propagation(\cite{defferrard2016convolutional,lingam2022piece}).

From this perspective, the aggregation step in Q-AGNN can be expressed as a second-order polynomial graph filter applied to the PQC-generated embeddings:
\begin{equation}
    H = \left( \beta_0 I + \beta_1 A + \beta_2 A^2 \right) Z,
\end{equation}

where $Z \in \mathbb{R}^{N \times d_q}$ is the matrix of quantum embeddings, and $\{\beta_k\}_{k=0}^2$ are implicit, data-dependent coefficients induced by the attention mechanism. Polynomial filters of this form are known to balance locality and contextual awareness while avoiding the excessive smoothing associated with deep message passing(\cite{oono2019graph,huang2024universal,li2024another}). By restricting aggregation to low-order neighborhoods, Q-AGNN preserves discriminative local structure while incorporating sufficient relational context for intrusion detection.

The attention mechanism further refines this filtering process by introducing adaptive, node-level weighting. Let $\alpha_v$ denote the learned attention weight associated with node $v$, normalized across the graph. These weights modulate the contribution of each node during aggregation, effectively reweighting the polynomial filter coefficients in a data-dependent manner. Unlike edge-wise attention mechanisms, this node-centric formulation reduces computational complexity and is well suited to PQC-based embeddings, where repeated quantum evaluations are costly. This design is particularly advantageous in similarity-based traffic graphs, where benign and malicious flows may occupy overlapping neighborhoods and selective emphasis is required.

Crucially, Q-AGNN adopts a hybrid quantum-classical decomposition that cleanly separates nonlinear feature transformation from relational reasoning. The PQC is evaluated once per node to generate expressive embeddings, while all neighborhood aggregation, attention weighting, and classification are performed classically. This separation avoids repeated quantum circuit evaluations during message passing and enables efficient end-to-end training using standard optimization methods on classical hardware, noisy simulators, or near-term quantum devices.

\subsection{Theoretical Foundations: Quantum Feature Maps and Expressivity}

The design of Q-AGNN is grounded in principles from quantum feature mapping and kernel theory, which provide insight into the representational advantages of parameterized quantum circuits (PQCs) over classical feature transformations.

\subsubsection{PQC as an Implicit Quantum Feature Map}

Let $x \in \mathbb{R}^d$ denote the classical feature vector associated with a network flow. The PQC used in Q-AGNN defines a nonlinear mapping

\begin{equation}
\Phi_{\theta} : \mathbb{R}^d \rightarrow \mathcal{H}, 
\quad x \mapsto |\psi(x;\theta)\rangle,
\end{equation}

where $\mathcal{H}$ is a Hilbert space of dimension $2^n$ for an $n$-qubit system. 
This mapping is realized via angle encoding followed by a parameterized EfficientSU2 ansatz.

Node embeddings are obtained through expectation values of as set of Pauli observables $\{P_k\}$:

\begin{equation}
z_k(x) = \langle \psi(x;\theta) | P_k | \psi(x;\theta) \rangle.
\end{equation}

These expectation values define coordinates of x under an implicit quantum feature map. The kernel function induced by this measurement strategy between two inputs $x$ and $x'$ is:

\begin{equation}
    K_\theta(x, x')
=
\sum_{k}
\langle \psi(x;\theta) \mid P_k \mid \psi(x;\theta) \rangle
\cdot
\langle \psi(x';\theta) \mid P_k \mid \psi(x';\theta) \rangle
\end{equation}

This is the Pauli expectation kernel, which measures similarity through the alignment of observable measurement outcomes in the quantum feature space. We note this is distinct from the fidelity kernel:
\begin{equation}
    K_{\theta}^{\text{fid}}(x, x')
=
\left|
\langle \psi(x;\theta) \mid \psi(x';\theta) \rangle
\right|^2
\end{equation}

which measures quantum state overlap directly. While both kernels are positive semi-definite and theoretically valid, Q-AGNN implicitly operates under the Pauli expectation kernel, since node embeddings are derived from observable measurements rather than state inner products. The two kernels coincide only in special cases, such as when the observable set ${P_k}$ forms a complete basis for the operator space, i.e., when tomographically complete measurements are performed. In the general case with a finite, fixed set of Pauli observables, the Pauli expectation kernel captures a projected view of the full quantum state geometry, sufficient for learning discriminative representations while remaining computationally tractable on near-term hardware.

\subsubsection{Expressivity Compared to Classical MLPs}

A classical multilayer perceptron (MLP) with ReLU or sigmoid activations implements piecewise-linear function approximation whose expressivity grows polynomially with width and depth.

In contrast, PQCs composed of parameterized rotations and entangling gates generate functions of the form

\begin{equation}
f(x) = 
\langle 0 | U^\dagger(x,\theta) \hat{O} U(x,\theta) | 0 \rangle,
\end{equation}

which, following Schuld et al.(\cite{schuld2021effect}), can be expressed as truncated Fourier series over the input features:
\begin{equation}
    f(x) = \sum_{\omega \in \Omega} c_{\omega} \, e^{i \, \omega \cdot x},
\end{equation}

where $\omega$ denotes the set of accessible frequency components and $c_{\omega}$ are complex coefficients determined by the trainable parameters $\theta$.

Critically, the accessible frequency spectrum $\Omega$ is not arbitrary, it is determined by the eigenvalue spectrum of the encoding Hamiltonian. Under $R_Y$ angle encoding, each qubit contributes a generator with eigenvalues $\{-\tfrac{1}{2}, +\tfrac{1}{2}\}$, yielding a frequency spectrum bounded by
\begin{equation}
    \Omega \subseteq \left\{-\frac{n}{2}, \ldots, 0, \ldots, \frac{n}{2}\right\},
\end{equation}

for an $n$-qubit system. In Q-AGNN, with $n = 4$ qubits, the maximum accessible frequency per input feature is therefore $\pm 2$, corresponding to second-order trigonometric interactions of the form
\begin{equation}
    \{1, \sin(x_i), \cos(x_i), \sin(2x_i), \cos(2x_i)\}.
\end{equation}

Cross-feature interactions arise through entangling CNOT gates, which introduce multiplicative coupling between qubits, enabling terms such as
\begin{equation}
    \sin(x_i)\cos(x_j), \quad \cos(x_i)\cos(x_j).
\end{equation}

However, the total number of accessible Fourier coefficients grows as $O(n^2)$ under linear entangling topology rather than exponentially, due to the constrained connectivity of the \emph{EfficientSU2} ansatz employed. This represents a meaningful but bounded expressivity advantage over a shallow classical MLP of equivalent parameter count, rather than an unrestricted exponential advantage.
Furthermore, the concentration of measure phenomenon implies that in high-dimensional quantum systems, random parameter initializations tend to produce expectation values concentrated near zero, potentially leading to flat optimization landscapes (barren plateaus) during training. For the shallow, 4-qubit circuits used in Q-AGNN, this effect is limited but non-negligible and is partially mitigated by the hybrid classical optimization pipeline.

While PQCs introduce a structurally distinct inductive bias compared to classical MLPs by embedding feature interactions through unitary evolution rather than layered nonlinear compositions, the practical expressivity advantage in Q-AGNN is more accurately described as moderate and dependent on circuit depth rather than inherently superior. This consideration is especially relevant in the four-qubit shallow-circuit regime imposed by current NISQ hardware constraints. It also motivates future investigation of deeper parameterized circuits and alternative encoding strategies as fault-tolerant quantum hardware becomes available.

\subsubsection{Boundedness and Optimization Stability}

Expectation values of Pauli observables satisfy

\begin{equation}
z_k(x) \in [-1,1],
\end{equation}

ensuring that quantum embeddings remain uniformly bounded. 
This boundedness prevents uncontrolled activation growth and promotes numerical stability when embeddings are composed with classical graph filters.

Unlike classical deep networks, which may require explicit normalization strategies to mitigate exploding or vanishing activations, PQC-based embeddings inherently preserve bounded magnitudes due to the unitary nature of quantum evolution.

This boundedness property provides an additional practical advantage under realistic NISQ conditions. Depolarizing noise, which is the dominant error channel on current quantum hardware, drives quantum states toward the maximally mixed state and causes expectation values of Pauli observables to decay toward zero approximately by a factor of $(1 - p)^d$, where $p$ denotes the depolarizing error rate per gate and $d$ represents the circuit depth. Since PQC embeddings in Q-AGNN are inherently bounded within the interval $[-1,1]$, this noise-induced attenuation does not generate out-of-distribution outputs. Instead, it produces embeddings that remain within the same bounded range encountered during training, although with reduced magnitude. As a result, the downstream attention mechanism and polynomial graph filter receive inputs whose distributional support is preserved under noise, even when individual embedding magnitudes are attenuated. In contrast, classical deep networks with unbounded activations may exhibit large deviations under hardware-induced perturbations, as activation magnitudes are not inherently constrained. The empirical consistency observed between noiseless and noisy simulation results in Section~\ref{noisy} aligns with this theoretical expectation.

\subsubsection{Interaction with Polynomial Graph Filtering}

Q-AGNN applies a second-order polynomial graph filter over PQC embeddings:

\begin{equation}
H = (\beta_0 I + \beta_1 A + \beta_2 A^2) Z,
\end{equation}

where $Z$ contains the PQC-generated node embeddings.

From a kernel perspective, this corresponds to performing relational smoothing in a quantum-induced feature space. 
The architecture therefore decouples nonlinear feature transformation (quantum domain) from relational aggregation (classical graph filtering).

This separation avoids repeated quantum circuit evaluations during message passing while preserving expressive nonlinear embeddings. 
The resulting model can thus be interpreted as performing kernelized graph filtering with a learnable quantum feature map.

\subsubsection{Implications for Intrusion Detection}

Intrusion detection requires modeling nonlinear dependencies across traffic statistics and relational propagation across communication patterns. 
The PQC provides expressive nonlinear transformations capable of capturing higher-order correlations, while the polynomial graph filter propagates these representations across network structure.

Consequently, Q-AGNN combines:

\begin{itemize}
\item Exponential Hilbert-space embedding capacity,
\item Bounded expectation-value representations,
\item Efficient low-order graph signal propagation.
\end{itemize}

This theoretical foundation explains the empirical improvements observed across multiple datasets and clarifies the structural advantages of hybrid quantum-classical graph learning for cybersecurity applications.

\section{Experimental Setup}
This section presents a rigorous experimental framework and a set of diverse benchmark datasets used to comprehensively evaluate the effectiveness of the proposed Q-AGNN model. Within this framework, experiments are conducted on four representative network traffic datasets selected to capture both analytical complexity and practical deployment considerations.

The selected datasets span multiple levels of data granularity, ranging from feature-rich flow representations derived from packet captures to compact NetFlow summaries commonly adopted in operational intrusion detection systems. This design choice enables a comprehensive assessment of Q-AGNN’s ability to learn discriminative structural patterns while maintaining robustness and efficiency under realistic computational constraints.

\begin{itemize}

\item \textbf{BoT-IoT(\cite{koroniotis2019towards}):}
This dataset was generated in a controlled cyber range environment at UNSW Canberra to emulate realistic Internet-of-Things (IoT) botnet scenarios. The dataset combines benign traffic with diverse attack behaviors originating from compromised IoT devices. To facilitate experimental tractability, the dataset authors provide a representative subset comprising 5\% of the full dataset, along with the corresponding 10 most discriminative features, while preserving attack diversity. This author-provided subset is used in all experiments in this study.

\item \textbf{UNSW-NB15(\cite{moustafa2015unsw}):}
It is a large-scale benchmark dataset created using the IXIA PerfectStorm traffic generator. It consists of labeled network flow records characterized by rich statistical and protocol-level features. Its heterogeneity and complexity make it particularly suitable for evaluating Q-AGNN’s capacity to capture subtle attack signatures and complex inter-feature dependencies.

\item \textbf{NetFlow Variants (NF-BoT-IoT, NF-UNSW-NB15)(\cite{sarhan2020netflow}):}
 NetFlow-based abstraction of the original BoT-IoT and UNSW-NB15 datasets, where packet-level traffic is summarized into a compact set of canonical flow attributes. This representation reflects realistic monitoring conditions in operational networks and is used to assess the scalability and robustness of Q-AGNN under lightweight data representations.

\end{itemize}

Together, these four datasets provide a balanced and realistic evaluation setting, enabling a systematic examination of the performance, generalization capability, and practical applicability of Q-AGNN across heterogeneous traffic representations and deployment scenarios.

\subsection{Data Preprocessing}

Data preprocessing is a critical component of the experimental pipeline. All features containing missing values (NaNs) are removed to ensure data consistency and reliability. Categorical attributes are converted into numerical representations using the \texttt{LabelEncoder} provided by the \texttt{scikit-learn} library.

To enable execution on quantum simulators and mitigate computational overhead, traffic records are aggregated based on \emph{unique source IP-destination IP pairs}. For each such pair, numerical feature values are summarized using their mean, while the associated class label is determined by the most frequently occurring label within the group. This grouping step makes the data more manageable for quantum simulation while retaining the representative traffic behavior needed for evaluation.

After aggregation, all features are scaled using \emph{Min-Max normalization} to map values into a uniform range. Given the limitations of \emph{NISQ} devices and the exponential cost of quantum circuit simulation, \emph{Principal Component Analysis (PCA)} is subsequently applied to reduce the feature space to four dimensions, matching the practical input constraints of the quantum model.

The resulting dataset is then partitioned into 70\% training, 15\% validation, and 15\% testing splits to support robust model training, hyperparameter tuning, and unbiased performance evaluation. 

\subsection{Graph Construction}

The representation of network traffic as a graph is a critical component of the proposed Q-AGNN framework. In this approach, each node in the graph corresponds to a single network flow, defined by a unique source-destination IP address pair(\cite{hosler2024graph}).

Let $x_i \in \mathbb{R}^F$ denote the feature vector associated with node $i$. Edges between nodes are established based on feature-level similarity rather than physical proximity. Specifically, the cosine similarity between node feature vectors is computed, and a bi-directional edge is introduced whenever the similarity exceeds a threshold of $0.9$. Formally, an edge $e_{ij} \in E$ exists if

\begin{equation}
\frac{x_i. x_j}{\|x_i\| \, \|x_j\|} \geq 0.9.
\end{equation}

\begin{figure}[ht]
    \centering
    \includegraphics[width=0.30\linewidth]{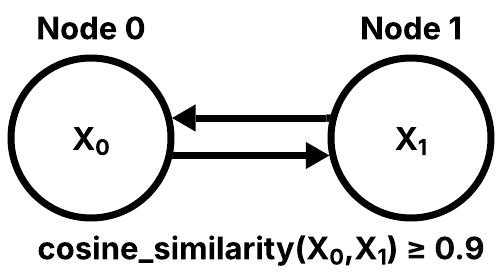}
    \caption{Illustration of the network flow graph construction. Each node represents a unique flow, and edges connect nodes with highly similar feature vectors.}
    \label{fig:graph}
\end{figure}

\begin{table}[t]
\centering
\caption{Graph statistics for training, validation, and testing sets.}
\label{tab:combined_graph_stats}
\small
\setlength{\tabcolsep}{4pt}  
\renewcommand{\arraystretch}{1.1}

\begin{tabular}{lcccccc}
\toprule
\multirow{2}{*}{\textbf{Dataset}} &
\multicolumn{2}{c}{\textbf{Training}} &
\multicolumn{2}{c}{\textbf{Validation}} &
\multicolumn{2}{c}{\textbf{Testing}} \\
\cmidrule(lr){2-3}
\cmidrule(lr){4-5}
\cmidrule(lr){6-7}
& \textbf{Nodes} & \textbf{Edges}
& \textbf{Nodes} & \textbf{Edges}
& \textbf{Nodes} & \textbf{Edges} \\
\midrule
BoT-IoT        & 109 & 1,392 & 23 & 66  & 24 & 40  \\
NF-BoT-IoT     & 126 & 5,870 & 27 & 296 & 28 & 324 \\
UNSW-NB15      & 217 & 11,086 & 47 & 392 & 47 & 578 \\
NF-UNSW-NB15   & 205 & 8,752 & 44 & 496 & 45 & 456 \\
\bottomrule
\end{tabular}
\end{table}

Fig.~\ref{fig:graph} presents a simple example of the resulting graph. Consider two network flows between IP address pairs \texttt{10.0.0.5}-\texttt{10.0.0.12} and \texttt{10.0.1.8}-\texttt{10.0.1.20}, represented by feature vectors $x_0$ and $x_1$, respectively. These flows are mapped to nodes in the graph, and a bi-directional edge is formed if their feature vectors exhibit sufficient cosine similarity. This allows the model to capture relationships between flows that share similar behavioral patterns, even if they originate from distinct endpoints.

This similarity-driven graph construction enables Q-AGNN to effectively model interactions among network flows, facilitating the learning of discriminative representations for intrusion detection. A summary of the resulting graph statistics for each dataset is provided in Table~\ref{tab:combined_graph_stats}.

\section{Training Configuration}

The proposed quantum-classical hybrid model was implemented using Python 3.11.7 with PyTorch 2.5.1+cu121, Qiskit 2.2.3, and Qiskit Machine Learning 0.9.0, enabling efficient GPU-accelerated computation and seamless integration of quantum neural network layers. Model training employed the Adam optimizer with a learning rate of 0.01 and an L2 regularization (weight decay) of $1\times10^{-2}$. The maximum number of epochs was set to 1000, and early stopping was applied if the validation loss did not improve for 30 consecutive epochs to prevent overfitting. During training, the Binary Cross-Entropy with Logits loss function was utilized to optimize the model for binary anomaly detection. For evaluation, the logits were passed through a sigmoid activation, with a threshold of 0.5 used to classify instances as anomalous or benign. 

Quantum computations were performed using different backends depending on the experiment. For noiseless simulations, the \texttt{StatevectorEstimator} was employed to obtain exact expectation values. To assess model performance under realistic noise conditions, an actual IBM Quantum backend was randomly selected from available operational devices for training. The noise profile from this backend was imported into the \texttt{AerSimulator} (using the density matrix method) to construct a noisy quantum simulator, which was then used with \texttt{BackendEstimatorV2} and a default precision of 0.03. For testing, the same backend and corresponding noise model were used to ensure consistency and reproducibility of the results. This setup enabled a controlled evaluation of the impact of quantum noise on model performance.  

For comparison, baseline models including GCN, GAT, GraphSAGE, ClusterGCN, GINConv, SuperGAT and TransformerConv were trained under identical conditions. Each baseline employed two hidden layers to capture two-hop neighborhood information, and the hidden layer dimensions were carefully selected such that the total number of trainable parameters closely matched that of the proposed quantum-classical framework. This ensured a fair and consistent evaluation across all models.

\section{Results and Discussion}

In this section, we present a comprehensive evaluation of the proposed Q-AGNN framework for network intrusion detection. The performance is assessed across multiple benchmark datasets, using a range of classical GNN baselines for comparison. We analyze the results under three key experimental settings: ideal, noise-free quantum estimation using the \texttt{StatevectorEstimator}, noisy quantum simulation with hardware-calibrated noise, and execution on real IBM quantum hardware.  


To evaluate the performance of Q-AGNN and for tractability, we adopt standard classification metrics commonly used in IDS. These include accuracy, precision, recall (detection rate), F1-score, false positive rate (FPR), false negative rate (FNR), and specificity, providing a comprehensive assessment of both detection effectiveness and error characteristics.




For precision, recall, and F1-score, we report macro-averaged values to equally consider both the normal and attack classes. This is particularly important in imbalanced intrusion detection datasets, where the attack class may be underrepresented. In contrast, FPR, FNR, and specificity are reported per class to provide operational insights, with a special focus on minimizing false positives to ensure practical applicability in real-world network monitoring.

\subsection{Embedding Space Analysis of PQC vs MLP}

\begin{figure}[t]
\centering

\begin{subfigure}{0.23\textwidth}
    \includegraphics[width=\linewidth]{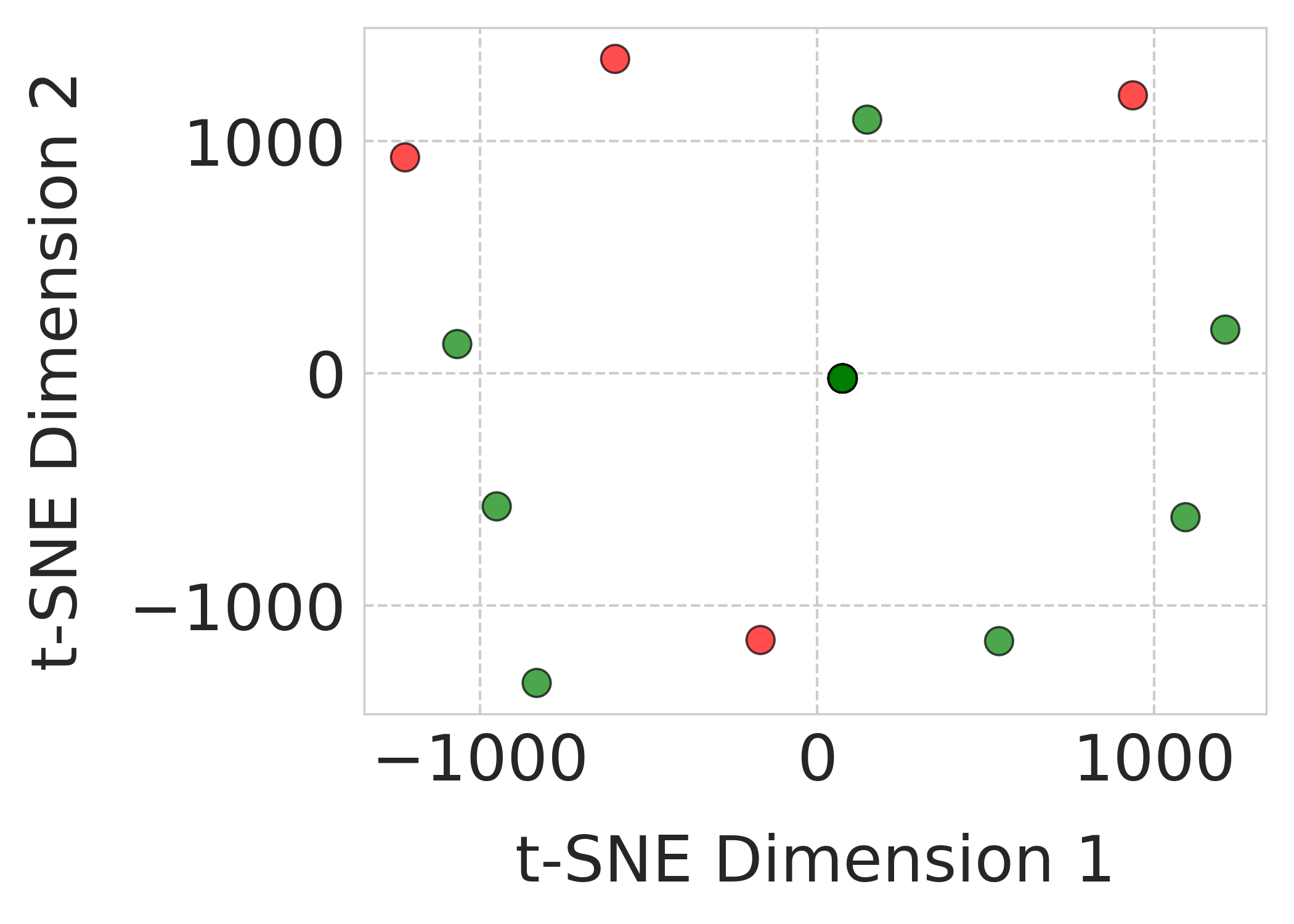}
    \caption{}
\end{subfigure}
\hfill
\begin{subfigure}{0.23\textwidth}
    \includegraphics[width=\linewidth]{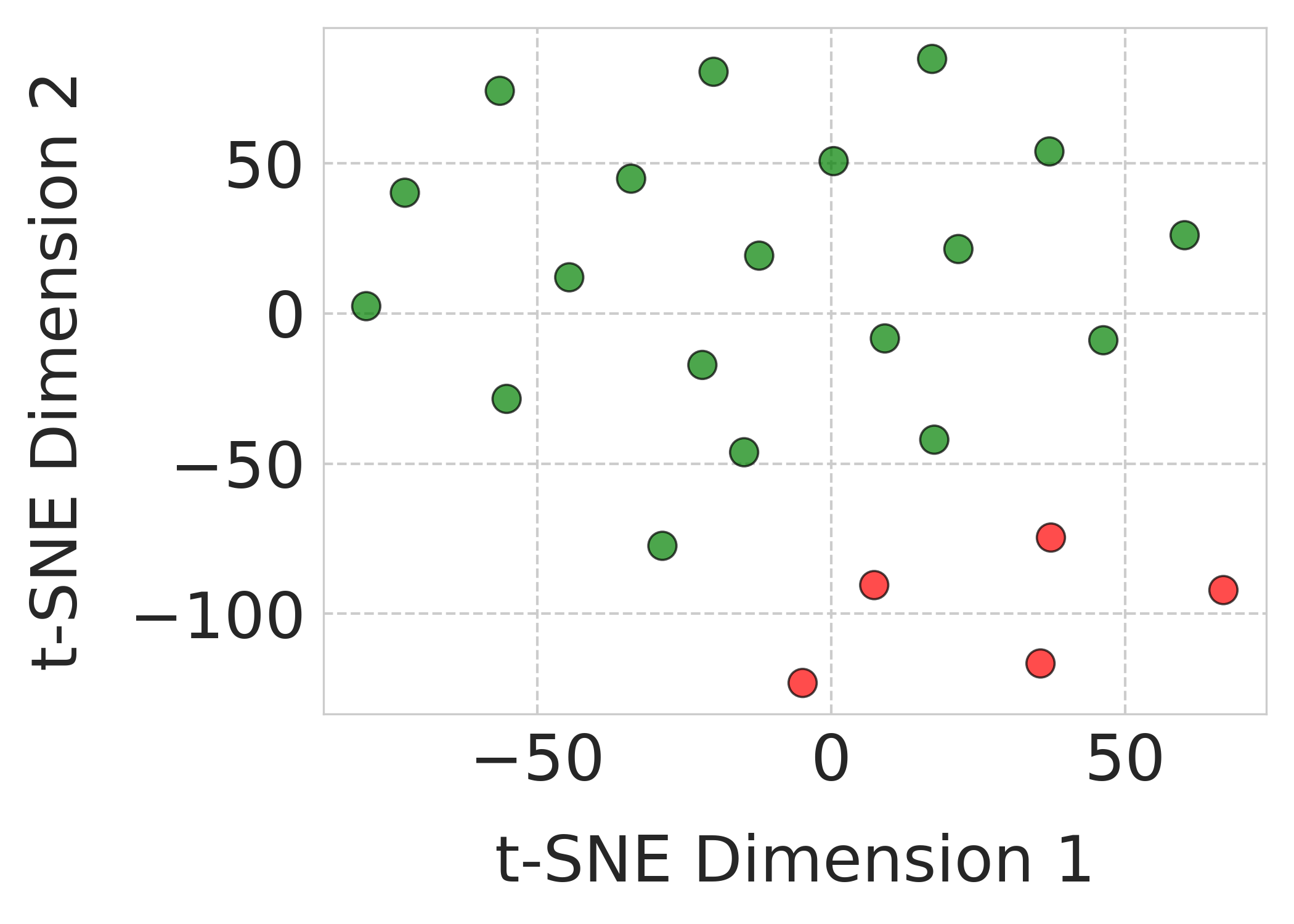}
    \caption{}
\end{subfigure}
\hfill
\begin{subfigure}{0.23\textwidth}
    \includegraphics[width=\linewidth]{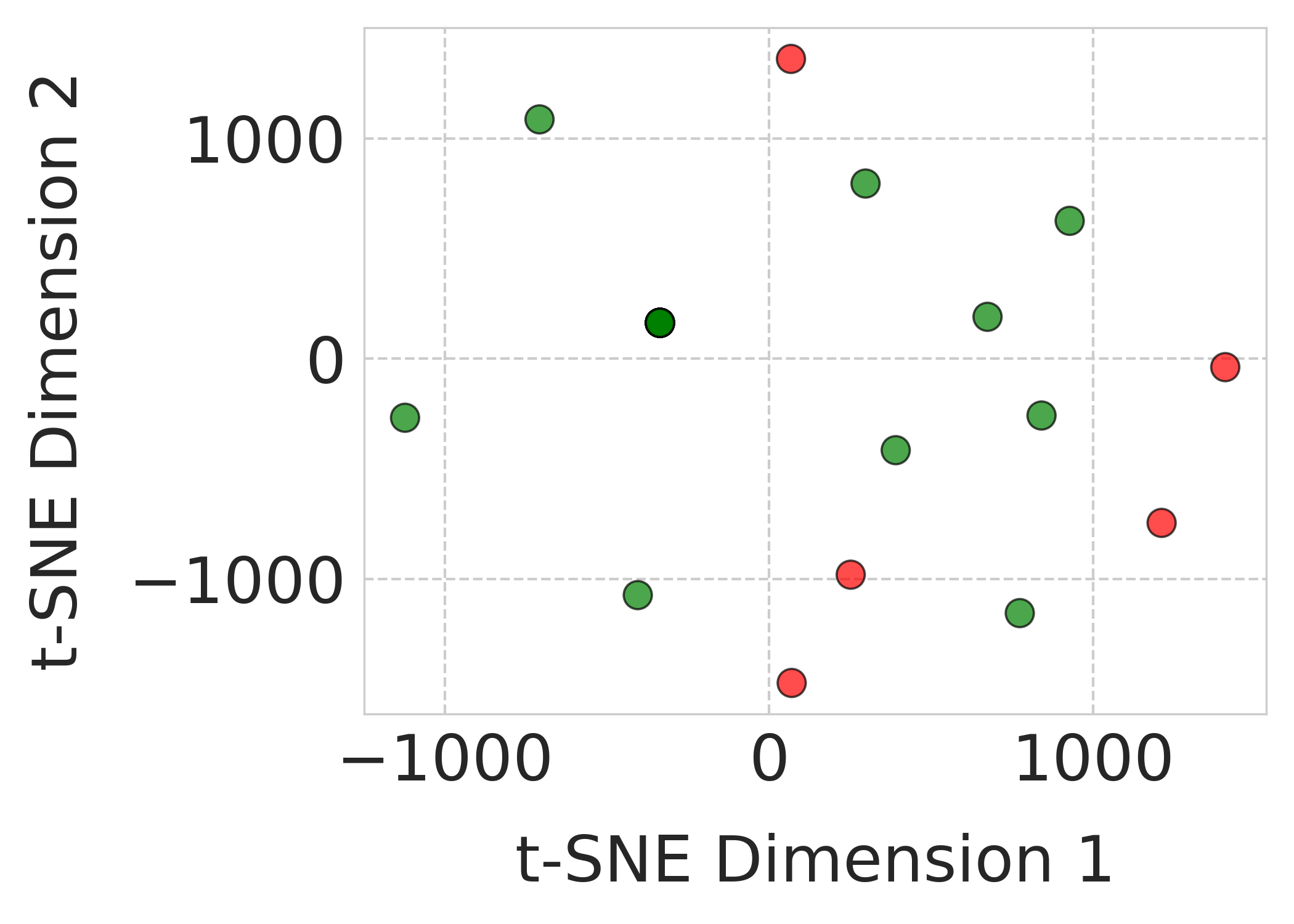}
    \caption{}
\end{subfigure}
\hfill
\begin{subfigure}{0.23\textwidth}
    \includegraphics[width=\linewidth]{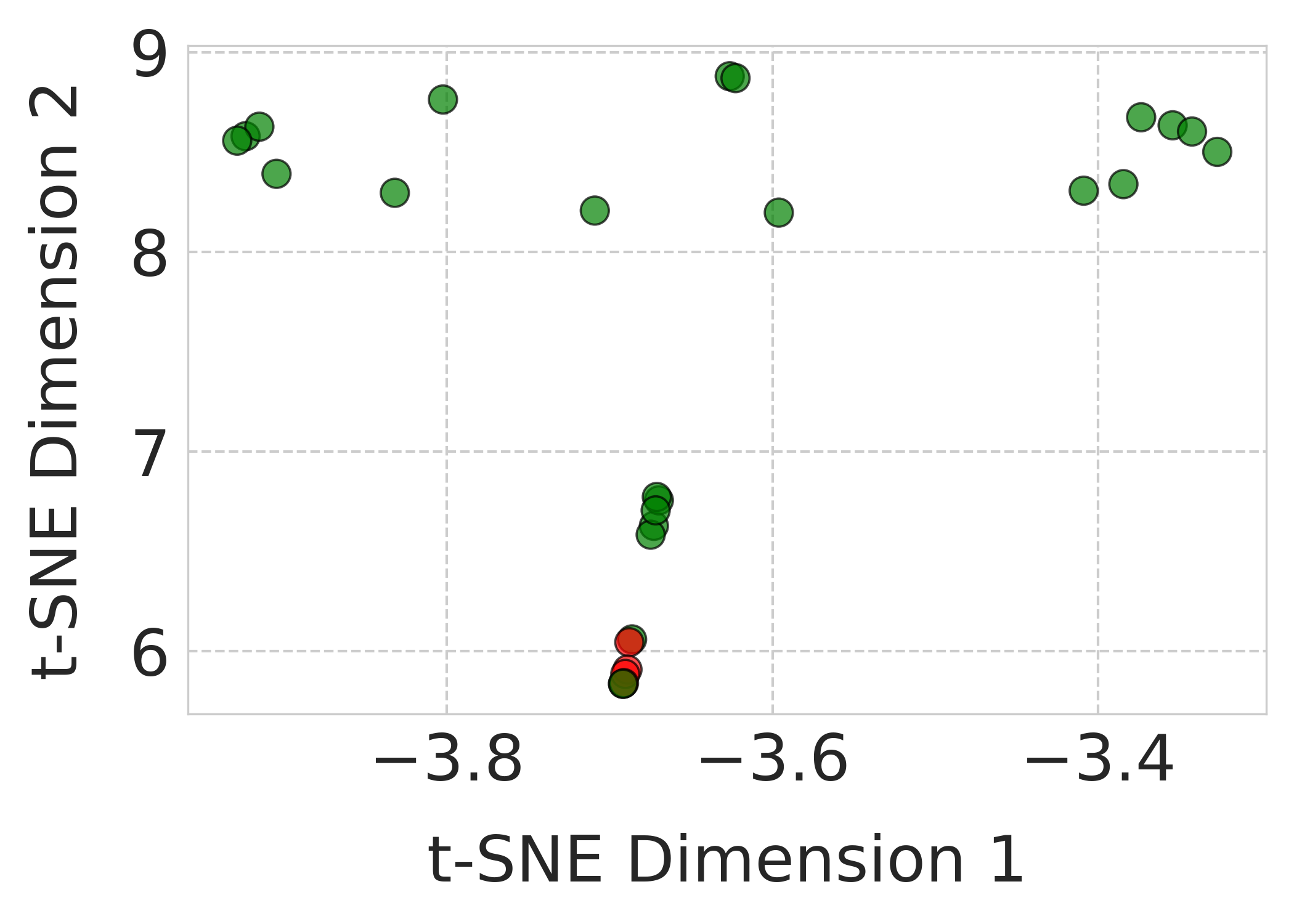}
    \caption{}
\end{subfigure}

\vspace{6pt}

\begin{subfigure}{0.23\textwidth}
    \includegraphics[width=\linewidth]{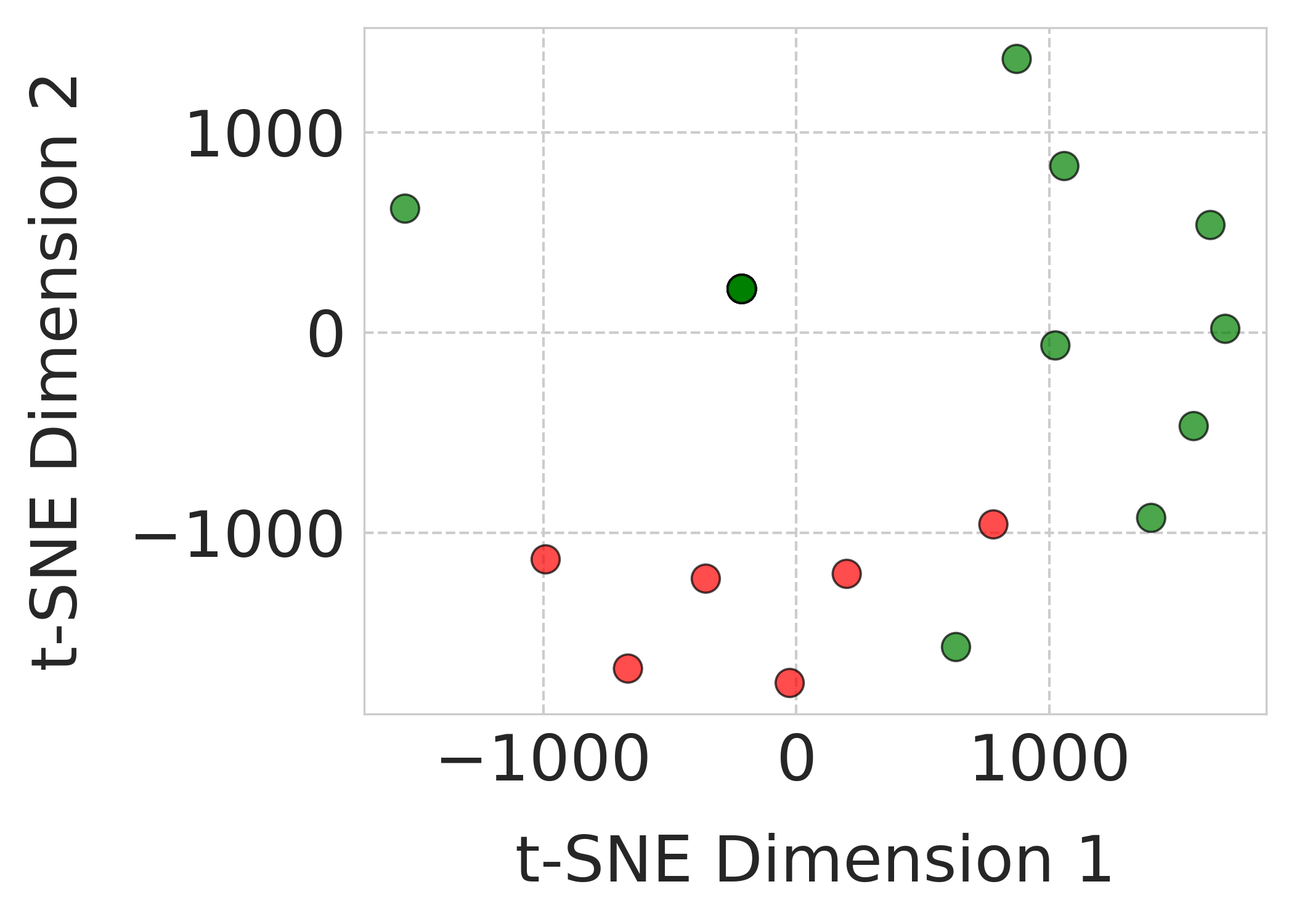}
    \caption{}
\end{subfigure}
\hfill
\begin{subfigure}{0.23\textwidth}
    \includegraphics[width=\linewidth]{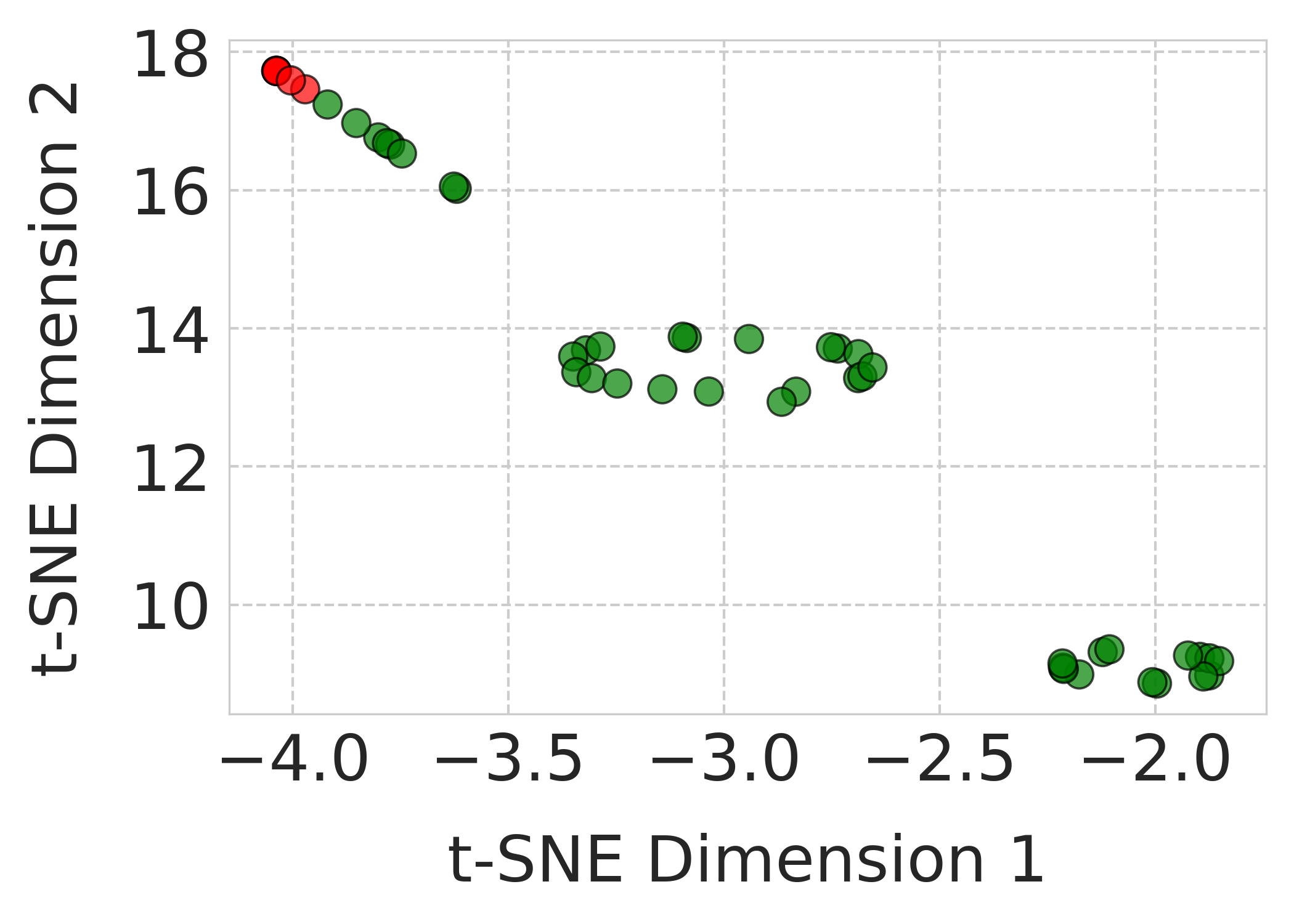}
    \caption{}
\end{subfigure}
\hfill
\begin{subfigure}{0.23\textwidth}
    \includegraphics[width=\linewidth]{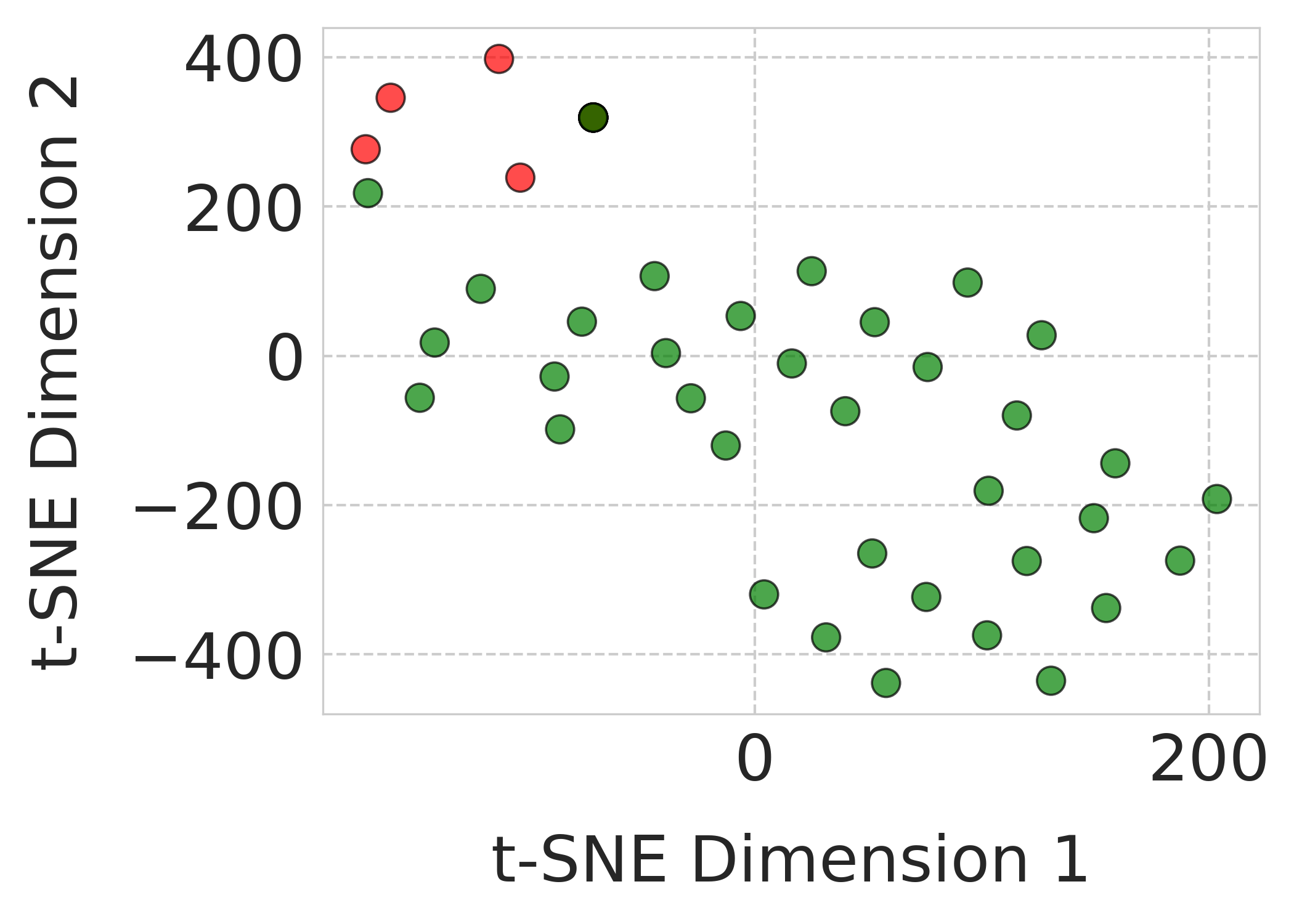}
    \caption{}
\end{subfigure}
\hfill
\begin{subfigure}{0.23\textwidth}
    \includegraphics[width=\linewidth]{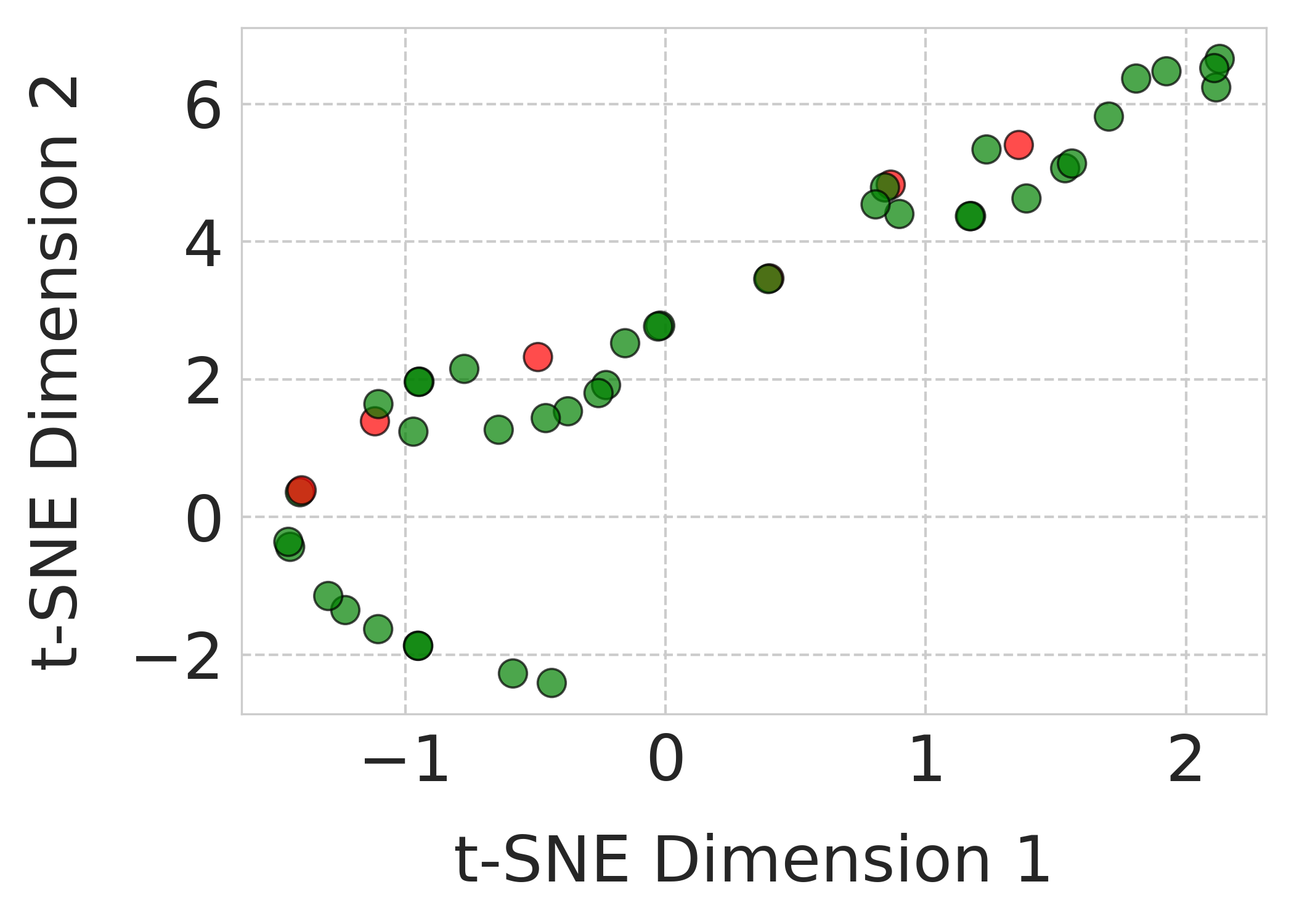}
    \caption{}
\end{subfigure}

\caption{
t-SNE visualization comparing embeddings produced by MLP and PQC.
(a)-(b): BoT-IoT dataset (MLP vs PQC).
(c)-(d): NF-BoT-IoT dataset (MLP vs PQC).
(e)-(f): UNSW-NB15 dataset (MLP vs PQC).
(g)-(h): NF-UNSW-NB15 dataset (MLP vs PQC).
Green points represent normal traffic and red points represent attack traffic.
}
\label{fig:tsne_mlp_pqc}
\end{figure}

Before evaluating performance across different quantum execution settings, we first analyze the geometry of the node embeddings learned by the model. Fig.~\ref{fig:tsne_mlp_pqc} presents t-SNE visualizations of these embeddings across all datasets for both the classical MLP-based model and the PQC-based model.

A consistent contrast is observed between the classical MLP and the PQC embeddings. The embeddings produced by the MLP are widely dispersed, with red and green points scattered across the space and frequently overlapping. The distribution appears noisy and unstructured, with no clear formation of compact regions corresponding to a specific class. In many areas, samples from different classes occupy the same neighborhoods, indicating that the MLP does not learn a representation space where similar behaviors are grouped together. Instead, it leaves much of the burden of discrimination to the final classifier layer rather than embedding the separability directly into the feature geometry.

In contrast, the PQC embeddings exhibit a markedly different spatial organization. Points belonging to the same class form visibly compact clusters, while samples from different classes are pushed into distinct regions of the space with noticeable gaps between them. The clusters are not only tighter but also more geometrically coherent, suggesting that the PQC transformation reshapes the feature space in a way that naturally groups similar traffic patterns together. This indicates that class separability is already encoded at the representation level, making the subsequent classification task significantly easier.

Another notable difference is the reduction in embedding spread. While MLP embeddings are stretched across large regions of the space, PQC embeddings occupy more confined and structured areas. This compactness reflects a form of implicit regularization, where the PQC discourages noisy or redundant representations and instead produces well-organized manifolds corresponding to underlying traffic behaviors.

\subsection{Sensitivity to Cosine Similarity Threshold}

\begin{table}[t]
\centering
\caption{Sensitivity of Q-AGNN to cosine similarity threshold on NF-UNSW-NB15.}
\label{tab:threshold_sensitivity}
\small
\setlength{\tabcolsep}{6pt}
\renewcommand{\arraystretch}{1.1}

\begin{tabular}{ccc}
\toprule
\textbf{Threshold} & \textbf{F1-score} & \textbf{FPR} \\
\midrule
0.60 & 0.816 & 0.000 \\
0.70 & 0.949 & 0.000 \\
0.80 & 1.000 & 0.000 \\
0.90 & 0.949 & 0.000 \\
0.95 & 0.949 & 0.000 \\
\bottomrule
\end{tabular}
\end{table}

As shown in Table~\ref{tab:threshold_sensitivity}, Q-AGNN exhibits stable performance across a broad range of cosine similarity thresholds. Increasing the threshold from 0.60 to 0.70 leads to a substantial improvement in F1-score, indicating that overly dense graphs introduce weak or noisy neighbor relationships that hinder effective aggregation. Thresholds between 0.70 and 0.95 yield consistently strong performance, with no observed increase in the false positive rate.

While the highest F1-score is achieved at a threshold of 0.80, we select 0.90 as the default value to favor sparser graphs with higher-confidence edges. This choice reduces graph density and computational overhead while maintaining stable detection performance, and is therefore better aligned with practical deployment considerations.

\subsection{Performance Using Statevector-Based Quantum Estimation}

\begin{table*}[t]
\centering
\caption{Performance of Q-AGNN, GCN, GAT, GraphSAGE, ClusterGCN, GINConv, SuperGAT, and TransformerConv across datasets. Precision, recall, and F1-score are reported as macro-averaged to equally account for both normal and attack classes. Accuracy, FPR, FNR, and specificity are reported as global or per-class metrics. Best metrics are highlighted in \textbf{bold}.}
\label{tab:results}

\resizebox{\textwidth}{!}{
\begin{tabular}{>{\centering\arraybackslash}m{3cm}lccccccc}
\toprule
\textbf{Datasets} & \textbf{Model} & \textbf{Acc} & \textbf{Prec} & \textbf{Rec} & \textbf{F1} & \textbf{FPR} & \textbf{FNR} & \textbf{Spec}\\
\midrule

\multirow{8}{*}{\textbf{BoT-IoT}}
& Q-AGNN (Ours)                              & \textbf{1.00} & \textbf{1.00} & \textbf{1.00} & \textbf{1.00} & \textbf{0.00} & \textbf{0.00} & \textbf{1.00} \\
& GCN                    & 0.9167 & 0.9524 & 0.8000 & 0.8500 & \textbf{0.00} & 0.4000 & \textbf{1.00} \\
& GAT           & 0.9167 & 0.9524 & 0.8000 & 0.8500 & \textbf{0.00} & 0.4000 & \textbf{1.00} \\
& GraphSAGE     & 0.9167 & 0.9524 & 0.8000 & 0.8500 & \textbf{0.00} & 0.4000 & \textbf{1.00} \\
& ClusterGCN        & 0.9167 & 0.9524 & 0.8000 & 0.8500 & \textbf{0.00} & 0.4000 & \textbf{1.00} \\
& GINConv             & 0.8333 & 0.7778 & 0.8947 & 0.7983 & 0.2105 & \textbf{0.00} & 0.7895 \\
& SuperGAT               & 0.9167 & 0.9524 & 0.8000 & 0.8500 & \textbf{0.00} & 0.4000 & \textbf{1.00} \\
& TransformerConv & 0.9167 & 0.9524 & 0.8000 & 0.8500 & \textbf{0.00} & 0.4000 & \textbf{1.00} \\
\midrule

\multirow{8}{*}{\textbf{NF-BoT-IoT}}
& Q-AGNN (Ours)                              & 0.8929 & 0.8106 & 0.8565 & 0.8303 & 0.0870 & 0.2000 & 0.9130 \\
& GCN                   & 0.8929 & 0.8106 & 0.8565 & 0.8303 & 0.0870 & 0.2000 & 0.9130 \\
& GAT           & 0.7857 & 0.4074 & 0.4783 & 0.4400 & \textbf{0.0435} & 1.0000 & \textbf{0.9565} \\
& GraphSAGE     & \textbf{0.9643} & \textbf{0.9167} & \textbf{0.9783} & \textbf{0.9434} & \textbf{0.0435} & \textbf{0.00} & \textbf{0.9565} \\
& ClusterGCN        & 0.9286 & 0.8571 & 0.9565 & 0.8939 & 0.0870 & \textbf{0.00} & 0.9130 \\
& GINConv             & 0.7500 & 0.7083 & 0.8478 & 0.7044 & 0.3043 & \textbf{0.00} & 0.6957 \\
& SuperGAT               & 0.8929 & 0.8333 & 0.7783 & 0.8014 & \textbf{0.0435} & 0.4000 & 0.9565 \\
& TransformerConv & 0.9286 & 0.8571 & 0.9565 & 0.8939 & 0.0870 & \textbf{0.00} & 0.9130 \\
\midrule

\multirow{8}{*}{\textbf{UNSW-NB15}}
& Q-AGNN (Ours)                              & \textbf{0.9787} & \textbf{0.9881} & \textbf{0.9167} & \textbf{0.9485} & \textbf{0.00} & \textbf{0.1667} & \textbf{1.00} \\
& GCN                   & \textbf{0.9787} & \textbf{0.9881} & \textbf{0.9167} & \textbf{0.9485} & \textbf{0.00} & \textbf{0.1667} & \textbf{1.00} \\
& GAT            & \textbf{0.9787} & \textbf{0.9881} & \textbf{0.9167} & \textbf{0.9485} & \textbf{0.00} & \textbf{0.1667} & \textbf{1.00} \\
& GraphSAGE   & \textbf{0.9787} & \textbf{0.9881} & 0.9161 & \textbf{0.9485} & \textbf{0.00} & \textbf{0.1667} & \textbf{1.00} \\
& ClusterGCN        & \textbf{0.9787} & \textbf{0.9881} & 0.9161 & \textbf{0.9485} & \textbf{0.00} & \textbf{0.1667} & \textbf{1.00} \\
& GINConv              & 0.8723 & 0.4362 & 0.5000 & 0.4659 & \textbf{0.00} & 1.0000 & \textbf{1.00} \\
& SuperGAT                & \textbf{0.9787} & \textbf{0.9881} & 0.9161 & \textbf{0.9485} & \textbf{0.00} & \textbf{0.1667} & \textbf{1.00} \\
& TransformerConv & 0.9574 & 0.9045 & 0.9045 & 0.9045 & 0.0244 & 0.1667 & 0.9756 \\
\midrule

\multirow{8}{*}{\textbf{NF-UNSW-NB15}}
& Q-AGNN (Ours)                              & \textbf{0.9111} & \textbf{0.9535} & \textbf{0.6667} & \textbf{0.7256} & \textbf{0.00} & \textbf{0.6667} & \textbf{1.00} \\
& GCN                 & 0.8667 & 0.4333 & 0.5000 & 0.4643 & \textbf{0.00} & 1.0000 & \textbf{1.00} \\
& GAT           & 0.8667 & 0.4333 & 0.5000 & 0.4643 & \textbf{0.00} & 1.0000 & \textbf{1.00} \\
& GraphSAGE     & 0.8667 & 0.4333 & 0.5000 & 0.4643 & \textbf{0.00} & 1.0000 & \textbf{1.00} \\
& ClusterGCN        & 0.8667 & 0.4333 & 0.5000 & 0.4643 & \textbf{0.00} & 1.0000 & \textbf{1.00} \\
& GINConv          & 0.1778 & 0.2857 & 0.1026 & 0.1509 & 0.7949 & 1.0000 & 0.2051 \\
& SuperGAT                & 0.8667 & 0.4333 & 0.5000 & 0.4643 & \textbf{0.00} & 1.0000 & \textbf{1.00} \\
& TransformerConv & 0.8667 & 0.4333 & 0.5000 & 0.4643 & \textbf{0.00} & 1.0000 & \textbf{1.00} \\
\bottomrule
\end{tabular}
}

\end{table*}

As a baseline, we first evaluated the Q-AGNN model under ideal, noise-free conditions using Qiskit's \texttt{StatevectorEstimator}. This estimator simulates the exact quantum state evolution without incorporating any hardware noise, providing an upper bound on the achievable model performance. 

Table~\ref{tab:results} compares Q-AGNN with several classical GNN baselines across all datasets in this setting. The results show that Q-AGNN is highly competitive overall, achieving the best or tied-best performance on BoT-IoT, UNSW-NB15, and NF-UNSW-NB15. In particular, it attains perfect performance on BoT-IoT and strong results on UNSW-NB15, indicating that the proposed architecture can effectively separate benign and malicious traffic under ideal quantum estimation.

For the more challenging NetFlow-based datasets, the results are more nuanced. On NF-BoT-IoT, GraphSAGE achieves the strongest overall performance, exceeding Q-AGNN in both macro-F1 and accuracy, while also obtaining a lower FPR. This suggests that on some datasets, well-established classical GNNs can remain highly competitive, or even superior, under the current experimental setting. However, on NF-UNSW-NB15, Q-AGNN yields the best overall balance among the reported metrics. Although several classical baselines achieve the same FPR as Q-AGNN, they do so with substantially worse recall and FNR, indicating that they miss a large fraction of attack instances. In contrast, Q-AGNN preserves zero FPR while improving recall and macro-F1, which is more desirable in intrusion detection scenarios where both false alarms and missed attacks are important.

The strong performance of Q-AGNN on several datasets, particularly NF-UNSW-NB15, may be attributed to its use of PQC-based feature encoding, which can capture higher-order correlations in network traffic graphs that may be difficult for classical GNNs to represent. When combined with graph aggregation, this representation appears to improve the balance between false positives and missed attacks, as reflected in the favorable recall and macro-F1 scores on the more challenging settings.

\begin{figure}[htbp]
    \centering
    \subfloat[]{
        \includegraphics[width=0.46\linewidth]{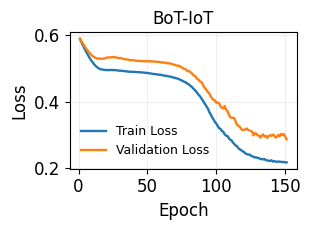}
        \label{fig:bot_iot_loss}
    }
    \hfill
    \subfloat[]{
        \includegraphics[width=0.46\linewidth]{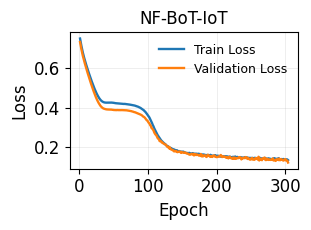}
        \label{fig:nfbot_loss}
    }

    \vspace{2mm} 

    \subfloat[]{
        \includegraphics[width=0.46\linewidth]{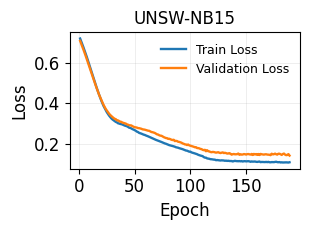}
        \label{fig:unsw_loss}
    }
    \hfill
    \subfloat[]{
        \includegraphics[width=0.46\linewidth]{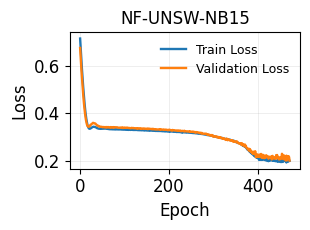}
        \label{fig:nfunsw_loss}
    }

    \caption{Train and validation loss curves for four datasets: 
    (a) BoT-IoT, 
    (b) NF-BoT-IoT, 
    (c) UNSW-NB15, 
    (d) NF-UNSW-NB15.}
    \label{fig:loss_curves_2x2}
\end{figure}

Fig.~\ref{fig:loss_curves_2x2} illustrates the training and validation loss dynamics of Q-AGNN across four datasets with varying graph characteristics. Distinct convergence behaviors are observed: rapid and smooth convergence for structurally cleaner graphs (NF-BoT-IoT and UNSW-NB15), slower stabilization for noisier IoT traffic graphs (BoT-IoT), and notably delayed improvement for the sparse and challenging NF-UNSW-NB15 graph. This variation indicates that the model adapts to the underlying graph complexity rather than memorizing features. Importantly, in all cases the validation loss closely follows the training loss, demonstrating stable generalization and the absence of overfitting while learning higher-order neighborhood representations.

These results establish a strong performance benchmark in an ideal, noise-free setting. Importantly, they also indicate the potential benefits of harnessing the full capacity of quantum computation. As fault-tolerant quantum hardware becomes available, larger graphs and more expressive quantum embeddings can be processed, potentially further improving both detection accuracy and robustness. In essence, the current results provide empirical evidence that the Q-AGNN framework is well-positioned to scale with future quantum computational resources while maintaining operationally meaningful low false positive rates.

\subsection{Performance Under Noisy Quantum Simulation with Hardware-Calibrated Noise}
\label{noisy}

To assess the robustness of the proposed Q-AGNN under realistic NISQ conditions, we evaluated the model on a noisy quantum simulator, where noise characteristics were imported from actual IBM quantum hardware. Table~\ref{tab:noisy_sim_results} summarizes the performance of the Q-AGNN model under noisy simulation settings across multiple datasets. 

\begin{table}[t]
\centering
\caption{Performance of Q-AGNN under noisy simulation with hardware-calibrated noise (HW). Precision, recall, and F1-score are reported as macro-averaged to equally account for both normal and attack classes. Accuracy, FPR, FNR, and specificity are reported as global or per-class metrics.}
\label{tab:noisy_sim_results}
\small
\setlength{\tabcolsep}{3pt}
\renewcommand{\arraystretch}{1.05}

\begin{tabular}{@{}l l c c c c c c c c@{}}
\toprule
\textbf{Model} & \textbf{Dataset} & Acc & Prec & Rec & F1 & FPR & FNR & Spec & HW \\
\midrule

\multirow{4}{*}{\textbf{Q-AGNN}}
& BoT-IoT      & 0.917 & 0.952 & 0.800 & 0.850 & 0.000 & 0.400 & 1.000 & Torino \\
& NF-BoT-IoT   & 0.929 & 0.857 & 0.957 & 0.894 & 0.087 & 0.000 & 0.913 & Torino \\
& UNSW-NB15    & 0.979 & 0.988 & 0.917 & 0.949 & 0.000 & 0.167 & 1.000 & Torino \\
& NF-UNSW-NB15 & 0.867 & 0.433 & 0.500 & 0.464 & 0.000 & 1.000 & 1.000 & Fez \\

\bottomrule
\end{tabular}
\end{table}

As observed, the Q-AGNN with imported hardware noise demonstrates performance largely consistent with the ideal StatevectorEstimator, achieving high accuracy, precision, and F1-score on the BoT-IoT and UNSW-NB15 datasets. Performance degradation is observed in certain datasets, particularly in the NF-UNSW-NB15 scenario, reflecting the impact of realistic quantum noise. Nevertheless, the results suggest that the proposed Q-AGNN framework remains feasible under hardware-calibrated noise conditions, although its robustness appears to be dataset-dependent and should be interpreted with caution when considering execution on actual quantum devices.

\subsection{Execution on Real Quantum Hardware}

To complement the simulation study, we executed the Q-AGNN framework on actual IBM quantum hardware with the sole objective of validating the practical operability of the proposed pipeline and examining whether the training dynamics observed in simulation persist under real NISQ conditions. This experiment is intended as a feasibility and behavioral validation, not as a performance evaluation.

\begin{figure}[htbp]
\centering
\begin{subfigure}[b]{0.47\linewidth}
\centering
\includegraphics[width=\linewidth]{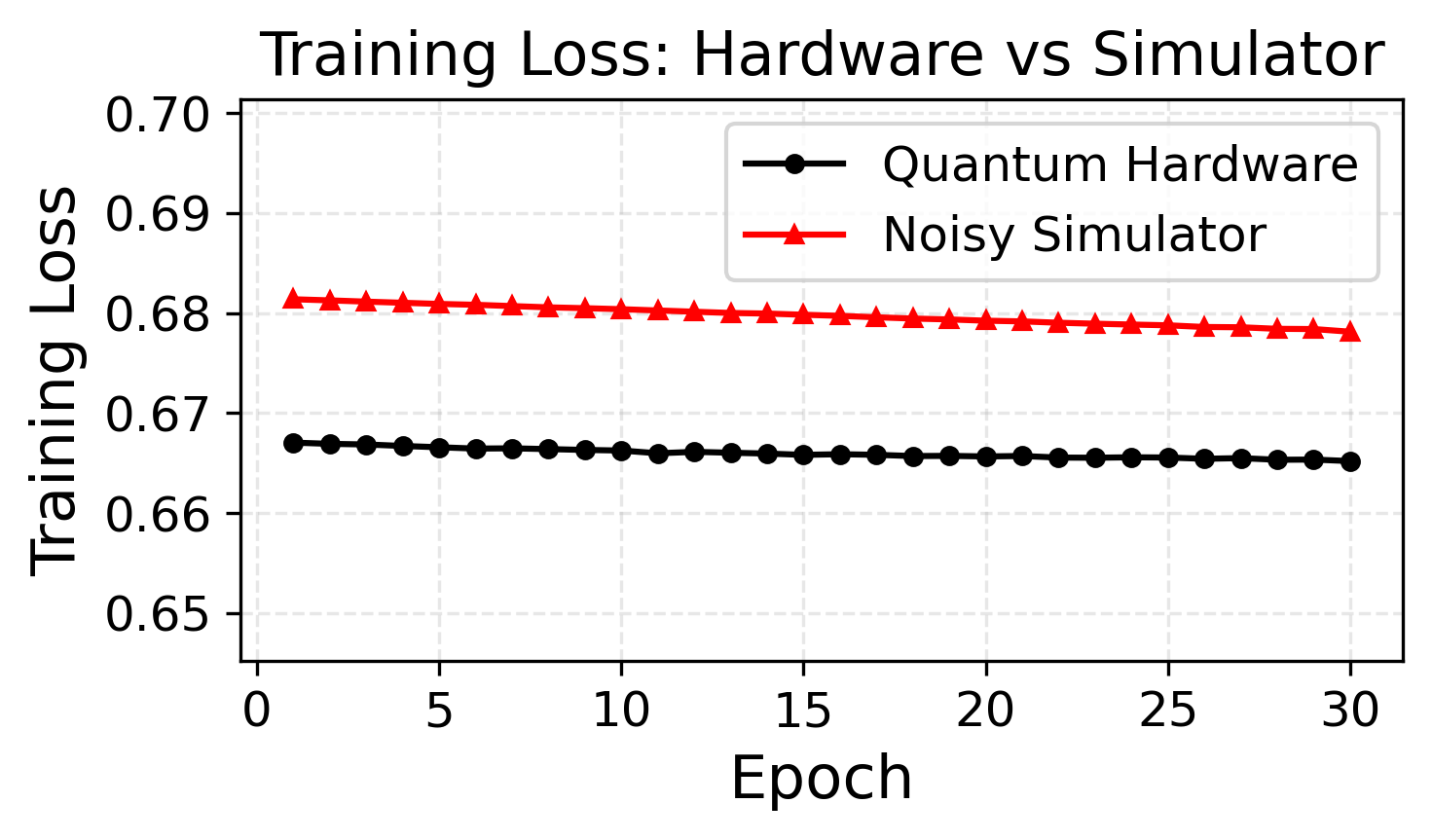}
\caption{}
\label{fig:train_loss_hw}
\end{subfigure}
\hfill
\begin{subfigure}[b]{0.47\linewidth}
\centering
\includegraphics[width=\linewidth]{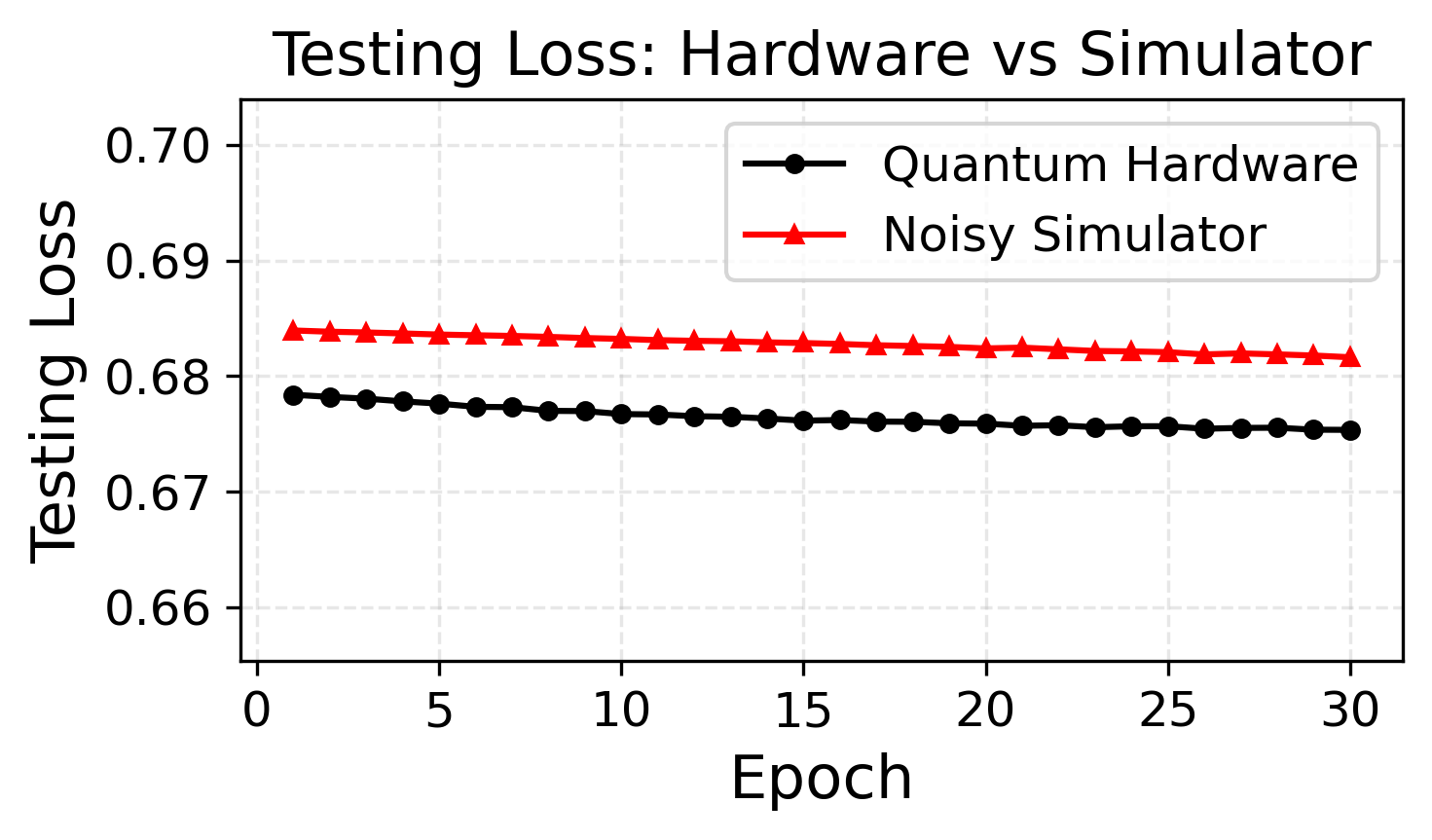}
\caption{}
\label{fig:test_loss_hw}
\end{subfigure}
\caption{Training and testing loss per epoch for Q-AGNN executed on IBM quantum hardware and on a noisy simulator using the same hardware-calibrated noise model. (a) Training loss. (b) Testing loss.}
\label{fig:hw_loss_curves}
\end{figure}

Given the limitations of current NISQ devices, including restricted qubit counts, queue times, and limited execution budgets, a reduced subset of the pre-processed UNSW-NB15 dataset was used. Specifically, 8 nodes were randomly sampled for training (5 normal flows and 3 attack flows) and 5 nodes for testing (3 normal flows and 2 attack flows). This minimal configuration ensured reliable execution on hardware while remaining sufficient to observe learning trends.

The quantum component was implemented using Qiskit’s SamplerQNN with 128 measurement shots to balance statistical stability and hardware cost. Due to execution-time constraints, training was performed for 30 epochs using the Adam optimizer with a learning rate of 0.001. These settings were selected to verify that the end-to-end Q-AGNN training process can be executed on real hardware and exhibit stable convergence behavior.

Experiments were run on the least busy available IBM backend (\texttt{ibm\_fez}) to reduce queue delays and temporal variations in noise characteristics. During execution, both training and testing loss values were recorded at each epoch to analyze convergence under realistic device noise.

For reference, the hardware-calibrated noise model of the same backend was imported into a noisy simulator, and the model was trained with identical data partitions and hyperparameters.

Fig.~\ref{fig:hw_loss_curves} shows that the loss trajectories obtained from real hardware closely follow those observed in the noisy simulator. The similarity in convergence behavior indicates that the simulator accurately reflects the device characteristics for this workload and, importantly, that the proposed Q-AGNN pipeline functions as intended on actual quantum hardware.

These results provide empirical evidence that Q-AGNN can be executed on current NISQ devices and that its learning dynamics remain consistent outside simulation, thereby validating the practical deployability of the framework under real hardware constraints.

\section{Ablation Study}

To assess the contribution of each component in Q-AGNN, we conduct an ablation study by selectively disabling attention, neighborhood depth, graph structure, or quantum encoding while keeping other components unchanged.

\begin{itemize}
    \item \textbf{PQC without Attention.} Retains quantum encoding and two-hop aggregation but removes attention, aggregating neighbors uniformly. This isolates the effect of attention.
    
    \item \textbf{One-Hop Neighborhood with Attention.} Considers only immediate neighbors with attention, disabling two-hop aggregation. This evaluates the benefit of higher-order neighborhoods.
    
    \item \textbf{One-Hop Neighborhood without Attention.} Restricts aggregation to one-hop neighbors and removes attention, assessing the combined impact of omitting higher-order dependencies and attention.
    
    \item \textbf{Q-AGNN without Graph Structure (Node-wise QNN).} Disables all message passing; each node is independently processed via the PQC. This isolates the contribution of quantum feature encoding without relational inductive bias.
    
    \item \textbf{MLP with Attention.} Replaces the PQC with a capacity-matched classical MLP, keeping attention and aggregation intact. This evaluates the gain from quantum encoding versus classical encoding.
    
    \item \textbf{MLP without Attention.} Combines a classical MLP with uniform aggregation (no attention), assessing whether performance gains stem from quantum encoding or classical graph aggregation alone.
\end{itemize}

\begin{table*}[t]
\centering
\caption{Comparison of Q-AGNN and ablation variants on BoT-IoT, NF-BoT-IoT, UNSW-NB15 and NF-UNSW-NB15 datasets. Precision, recall, and F1-score are reported as macro-averaged to equally account for both normal and attack classes. Accuracy, FPR, FNR, and specificity are reported as global or per-class metrics. Best results are highlighted in \textbf{bold}.}
\label{tab:ablation}

\resizebox{\textwidth}{!}{
\begin{tabular}{>{\centering\arraybackslash}m{2.6cm}lccccccc}
\toprule
\textbf{Dataset} & \textbf{Model} &
\textbf{Acc} &
\textbf{Prec} &
\textbf{Rec} &
\textbf{F1} &
\textbf{FPR} &
\textbf{FNR} &
\textbf{Spec} \\
\midrule

\multirow{7}{*}{BoT-IoT}
& Q-AGNN         & \textbf{1.00} & \textbf{1.00} & \textbf{1.00} & \textbf{1.00} & \textbf{0.00} & \textbf{0.00} & \textbf{1.00} \\
& PQC w/o Att    & 0.7917 & 0.3958 & 0.5000 & 0.4419 & \textbf{0.00} & 1.0000 & \textbf{1.00} \\
& 1-Hop + Att    & 0.9167 & 0.8737 & 0.8737 & 0.8737 & 0.0526 & 0.2000 & 0.9474 \\
& 1-Hop w/o Att  & 0.7917 & 0.3958 & 0.5000 & 0.4419 & \textbf{0.00} & 1.0000 & \textbf{1.00} \\
& Node-wise QNN  & 0.7917 & 0.3958 & 0.5000 & 0.4419 & \textbf{0.00} & 1.0000 & \textbf{1.00} \\
& MLP + Att      & 0.8750 & 0.8250 & 0.7737 & 0.7949 & 0.0526 & 0.4000 & 0.9474 \\
& MLP w/o Att    & 0.8750 & 0.8250 & 0.7737 & 0.7949 & 0.0526 & 0.4000 & 0.9474 \\
\midrule

\multirow{7}{*}{NF-BoT-IoT}
& Q-AGNN         & 0.8929 & 0.8106 & 0.8565 & 0.8303 & 0.0870 & 0.2000 & 0.9130 \\
& PQC w/o Att    & 0.8214 & 0.4107 & 0.5000 & 0.4510 & \textbf{0.00} & 1.0000 & \textbf{1.00} \\
& 1-Hop + Att    & 0.8929 & 0.8106 & 0.8565 & 0.8303 & 0.0870 & 0.2000 & 0.9130 \\
& 1-Hop w/o Att  & 0.8571 & \textbf{0.9259} & 0.6000 & 0.6267 & \textbf{0.00} & 0.8000 & \textbf{1.00} \\
& Node-wise QNN  & \textbf{0.9286} & 0.8571 & \textbf{0.9565} & \textbf{0.8939} & 0.0870 & \textbf{0.00} & 0.9130 \\
& MLP + Att      & 0.8929 & 0.8106 & 0.8565 & 0.8303 & 0.0870 & 0.2000 & 0.9130 \\
& MLP w/o Att    & 0.8214 & 0.4107 & 0.5000 & 0.4510 & \textbf{0.00} & 1.0000 & \textbf{1.00} \\
\midrule

\multirow{7}{*}{UNSW-NB15}
& Q-AGNN         & 0.9787 & 0.9881 & 0.9167 & 0.9485 & \textbf{0.00} & 0.1667 & \textbf{1.00} \\
& PQC w/o Att    & 0.8723 & 0.4362 & 0.5000 & 0.4659 & \textbf{0.00} & 1.0000 & \textbf{1.00} \\
& 1-Hop + Att    & \textbf{1.00} & \textbf{1.00} & \textbf{1.00} & \textbf{1.00} & \textbf{0.00} & \textbf{0.00} & \textbf{1.00} \\
& 1-Hop w/o Att  & 0.8723 & 0.4362 & 0.5000 & 0.4659 & \textbf{0.00} & 1.0000 & \textbf{1.00} \\
& Node-wise QNN  & \textbf{1.00} & \textbf{1.00} & \textbf{1.00} & \textbf{1.00} & \textbf{0.00} & \textbf{0.00} & \textbf{1.00} \\
& MLP + Att      & 0.9574 & 0.9045 & 0.9045 & 0.9045 & 0.0244 & 0.1667 & 0.9756 \\
& MLP w/o Att    & 0.8511 & 0.4348 & 0.4878 & 0.4598 & 0.0244 & 1.0000 & 0.9756 \\
\midrule

\multirow{7}{*}{NF-UNSW-NB15}
& Q-AGNN         & \textbf{0.91111} & \textbf{0.9535} & \textbf{0.6667} & \textbf{0.7256} & \textbf{0.00} & \textbf{0.6667} & \textbf{1.00} \\
& PQC w/o Att    & 0.8667 & 0.4333 & 0.5000 & 0.4643 & \textbf{0.00} & 1.0000 & \textbf{1.00} \\
& 1-Hop + Att    & 0.8667 & 0.4333 & 0.5000 & 0.4643 & \textbf{0.00} & 1.0000 & \textbf{1.00} \\
& 1-Hop w/o Att  & 0.8667 & 0.4333 & 0.5000 & 0.4643 & \textbf{0.00} & 1.0000 & \textbf{1.00} \\
& Node-wise QNN  & 0.8667 & 0.4333 & 0.5000 & 0.4643 & \textbf{0.00} & 1.0000 & \textbf{1.00} \\
& MLP + Att      & 0.8667 & 0.4333 & 0.5000 & 0.4643 & \textbf{0.00} & 1.0000 & \textbf{1.00} \\
& MLP w/o Att    & 0.8667 & 0.4333 & 0.5000 & 0.4643 & \textbf{0.00} & 1.0000 & \textbf{1.00} \\
\bottomrule
\end{tabular}
}
\end{table*}

Table~\ref{tab:ablation} presents the effect of the individual components in Q-AGNN. Attention emerges as the most consistently important factor. Removing attention often causes a sharp drop in macro-F1, and in several cases the resulting models exhibit near-degenerate behavior with very high specificity but very poor recall for the attack class. This suggests that uniform aggregation is often insufficient to highlight informative neighbors in these intrusion graphs.

Second, the effect of graph propagation is dataset-dependent. On some datasets, restricting the model to one-hop aggregation or even removing message passing entirely leads to only a limited degradation, and in a few cases node-wise or one-hop variants remain highly competitive. This suggests that for those datasets, node attributes already carry strong discriminative signals and higher-order relational information offers only marginal additional benefit. In contrast, on more challenging settings such as NF-UNSW-NB15, the full graph-based Q-AGNN substantially outperforms all reduced variants, showing that relational context becomes crucial when the detection task is harder.

Third, the role of the PQC should be interpreted jointly with the graph architecture rather than in isolation. While the PQC-based variants without attention are not consistently stronger than their classical MLP counterparts, the fairest comparison is between Q-AGNN and MLP with attention, where the aggregation mechanism is held fixed and only the encoder changes. Under this matched comparison, Q-AGNN achieves equal or better macro-F1 on all datasets, with especially clear gains on BoT-IoT, UNSW-NB15, and NF-UNSW-NB15. This suggests that the quantum encoder is most effective when coupled with attention-guided neighborhood aggregation, rather than as a standalone replacement for a classical feature encoder.

Overall, the ablation indicates that Q-AGNN benefits from the interaction of three elements, namely attention-based neighbor selection, graph-aware aggregation, and quantum feature encoding. Their contributions are not uniform across datasets, but the full combination yields the most robust overall behavior.

\section{Conclusion}

In this work, we propose Q-AGNN, a hybrid quantum-classical graph neural network for intrusion detection that combines classical message passing with PQC-based node embeddings to capture higher-order correlations in network graphs. Across multiple benchmark datasets, Q-AGNN achieves competitive or superior performance to classical GNN baselines while maintaining very low false positive rates, a critical requirement for practical IDS. Experiments with hardware-calibrated noisy simulations demonstrate robustness under realistic NISQ conditions, and execution on IBM quantum hardware confirms practical feasibility despite current device limitations. These results highlight the promise of hybrid quantum-classical graph learning for cybersecurity, with future work aimed at scaling PQC embeddings, enhancing noise mitigation, and exploring federated quantum learning for privacy-preserving intrusion detection.

\section*{Declarations}

\subsection*{Funding}
The authors wish to acknowledge that this research received no external funding or financial support.

\bibliography{ref}

\end{document}